\documentclass[prb,a4paper,twocolumn,superscriptaddress,nofootinbib]{revtex4-1}
\pdfoutput=1
\usepackage{graphicx} 
\usepackage{amsmath} \usepackage{array} \usepackage{amssymb}
\usepackage{amsfonts} \usepackage{color} \usepackage{enumitem}
\usepackage{bm} \usepackage{hyperref} \usepackage{verbatim}
\usepackage{morefloats} \usepackage{dcolumn} \usepackage{xspace}

\newcommand{\vektor}[1]{{\bf #1}}

 \newcommand{\ve}{\varepsilon}

\newcommand{\be}{\begin{equation*}} \newcommand{\ee}{\end{equation*}}
\newcommand{\bea}{\begin{eqnarray*}}
  \newcommand{\eea}{\end{eqnarray*}}

\newcommand{\specialcell}[2][c]{%
  \begin{tabular}[#1]{@{}c@{}}#2\end{tabular}}
\newcommand{\eref}[1]{Eq.~(\ref{#1})} \newcommand{\fref}[1]{Fig.~\ref{#1}} 
\newcommand{\sref}[1]{Sec.~\ref{#1}}
\newcommand{\tref}[1]{Table \ref{#1}}
\begin{document}

\title{Phase Diagram of the Spin-$1/2$ Triangular
    $J_1$-$J_2$ Heisenberg Model on a 3-leg Cylinder}

\author{S. N. Saadatmand} \email{s.saadatmand@uq.edu.au}
\affiliation{ARC Centre for Engineered Quantum Systems, School of Mathematics and Physics,
  The University of Queensland,
  St Lucia, QLD 4072, Australia}

\author{B. J. Powell}
\affiliation{School of Mathematics and Physics, The University of Queensland, 
  St Lucia, QLD 4072, Australia}

\author{I. P. McCulloch}
\affiliation{ARC Centre for Engineered Quantum Systems, School of Mathematics and Physics, 
  The University of Queensland,
  St Lucia, QLD 4072, Australia}

\begin{abstract}

  We study the phase diagram of the frustrated Heisenberg model on the triangular lattice
  with nearest and next-nearest neighbor spin exchange coupling, on 3-leg ladders.
  Using the density-matrix renormalization-group method, we obtain the complete
  phase diagram of the model, which includes quasi-long-range $120^\circ$ and 
  columnar order, and a Majumdar-Ghosh phase with short-ranged correlations. 
  All these phases are non-chiral and planar. We also identify the nature of phase transitions.

\end{abstract}



\date{\today}

\maketitle

\section{Introduction\label{sec:intro}}

Quantum magnetism in reduced dimensions gives rise to a fascinating range of 
behaviors.\cite{AuerbachBook,Sachdev11,Lhuillier08} In one-dimensional (1D) 
systems, powerful analytical and numerical methods have allowed a deep understanding 
of phenomena such as fractionalization,\cite{Tsvelik03} dimerization,\cite{MajumdarGhosh} 
and symmetry protected topological order.\cite{AKLT,Pollmann12} In two dimensions (2D) 
there remain many more open problems such as understanding spin 
liquids,\cite{Balent10,Powell11} intrinsic topological order,\cite{Chen10,Qi11} 
and the connection between exotic magnetic phases and 
unconventional superconductivity.\cite{Qiu11,Powell11} 
Few-leg ladders are a vital intermediate class as they
allow for the application of accurate numerical
methods\cite{Pang11} available for large 1D systems, while also providing important insights into 
new physics occurring in the crossover to two dimensions.\cite{Dagotto99,Kim08}

In an unfrustrated system, such as the nearest neighbor Heisenberg model on 
the square lattice, all of the terms in the Hamiltonian can be minimized 
simultaneously. This tends to favor long-range order in 2D. Therefore, 
frustrated systems are excellent candidates in which to search for  
exotic phases of matter without conventional ordering.\cite{Balent10,Powell11} 

The spin-$1/2$ triangular Heisenberg model (THM) is a prototypical
model for frustrated magnets in two dimension.\cite{Lhuillier08}
In 1973, Anderson\cite{Anderson73} suggested that the resonating-valence-bond
(RVB) state could play a pivotal role in the description of novel
magnetic materials, and his conjecture that ground-state of the spin-$1/2$ THM would be
an RVB state provoked much interest.
However studies of this model have failed to find
an RVB state and the evidence\cite{Jolicoeur89,Jolicoeur90} 
is now very strong that for the pure isotropic model 
with nearest-neighbor (NN) interactions, 
the ground-state is a
$120^{\circ}$ magnetically ordered state.\cite{Sachdev11}
Variants of the THM describe
some properties of organic materials\cite{Powell11,Balent10} such as
$\kappa$-(BEDT-TTF)$_2$Cu$_2$(CN)$_3$, EtMe$_3$Sb[Pd(dmit)$_2$]$_2$, 
EtMe$_3$P[Pd(dmit)$_2$]$_2$, Mo$_3$S$_7$(dmit)$_3$ \cite{Janani14,Jacko15} 
and also some quasi-two-dimensional inorganic
materials\cite{Svistov03,Zvyagin14,Starykh14,Powell11,Balent10,ColdeaMaterialSL} 
such as RbFe(MoO$_4$)$_2$, Ba$_3$CoSb$_2$O$_9$, Cs$_2$CuBr$_4$, and Cs$_2$CuCl$_4$.

In 1D the prototypical frustrated system is 
the zig-zag chain, which has
an exact solution at the Majumdar-Ghosh point\cite{MajumdarGhosh,AuerbachBook}
where the NN coupling is twice the next-nearest-neighbor (NNN)
coupling.
The ground-state is characterized by long-range dimer order and is two-fold degenerate.
As we show below, an NNN Majumdar-Ghosh phase is stabilized in a large region
of the phase diagram of the 3-leg triangular ladder.

So far the THM has been mostly considered with only NN exchange coupling, 
but additional interactions or anisotropies may stabilize 
exotic states. A
natural choice for an additional interaction, while retaining isotropy,
is an NNN coupling to add further frustration effects.
In this paper we study the $J_1$-$J_2$ THM 
on a width 3 cylinder as a simplified version of the full
2D model, but readily accessible to numerical methods. 
The ladder model has clear connections to the 2D THM and
also extrapolates smoothly to the Majumdar-Ghosh 
point of the zig-zag chain.
The $J_1$-$J_2$ THM in 2D has been previously studied using
semi-classical spin-wave theories (SWT) and exact
diagonalization,\cite{Dagotto89,Hirsch89,Bernu94,Jolicoeur90,Chubukov92,Deutscher93}
but these studies did not cover the physics of the whole phase
diagram. Recently, a coupled cluster study\cite{CCMSpinLiquid}
and quantum Monte Carlo (QMC) studies\cite{QMCSpinLiquid1,QMCSpinLiquid2} have identified
a phase in this model that is a candidate for a spin liquid.
Magnetically ordered states in a variety of classical $O(3)$ models with
$J_1$-$J_2$-$J_3$ interactions have been studied by 
Messio, Lhuillier, and Misguich,\cite{Messio11}
finding several different `regular magnetic orders' relevant to the triangular lattice,
including planar and non-planar $120^\circ$ states.

\begin{figure}
  \begin{center}
    \includegraphics[width=\columnwidth]{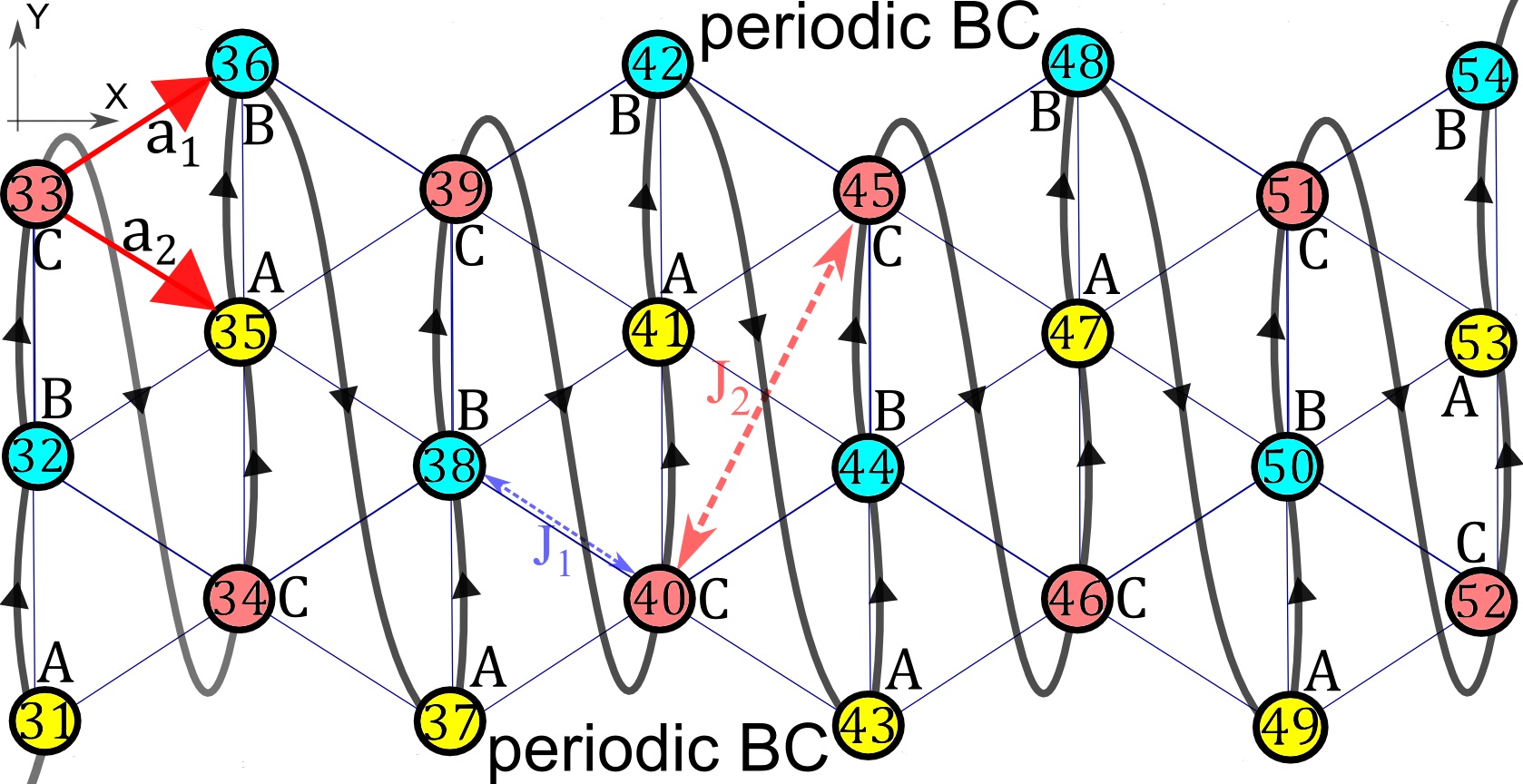}
    \caption{(Color online) Lattice structure and mapping of the 3-leg cylinder to
      the one-dimensional chain employed by the finite DMRG calculations. 
      Spins sit on vertices. The
      lattice is tripartite and the sub-lattices are labeled A, B, and
      C. ${\bf a_1}$ and ${\bf a_2}$ are lattice
      vectors.\label{fig:lattice}}
  \end{center}
\end{figure}

The lattice we consider is shown in \fref{fig:lattice}. The
Hamiltonian is
\begin{equation}
  \label{eq:Hamiltonian}
  H = J_1 \sum_{\langle i,j \rangle} \vektor{S}_i . {\bf S}_j + 
   J_2 \sum_{\langle\langle i,j \rangle\rangle} {\bf S}_i . {\bf S}_j \; ,
\end{equation}
where $\langle i,j \rangle$ ($\langle\langle i,j \rangle\rangle$) indicates 
that the sum is over all NN (NNN)
couplings. To cover the full range of couplings, we
introduce the following parameterisation:
\begin{equation}
  J_1 = J\cos\theta,~J_2 = J\sin\theta \; .
\end{equation}
Where the $J$ is the unit of energy and henceforth we fix
$J=1$. The main difficulty in studying the lattice shown in
\fref{fig:lattice} is frustration. The lowest energy state of the AFM
Heisenberg model on a square lattice has N\'{e}el
order.\cite{Lhuillier08} This cannot be formed on
an equilateral triangular lattice, and as a result there is competition
between terms in the Hamiltonian, \eref{eq:Hamiltonian}, and
they cannot simultaneously minimize their local energy.
Therefore it is clear that the lowest energy state must
be a compromise, such as the $120^{\circ}$ state. The $120^{\circ}$ 
state on the triangular lattice is less stable
than the  N\'{e}el state on the square lattice,\cite{Jolicoeur90,Powell11}
as the sublattice magnetization of the triangular lattice is significantly reduced
compared to its classical value. Because of this reduced stability inherent to the triangular
lattice, upon perturbing the Hamiltonian one may expect to see a variety of new phases.


There have been several numerical studies of the THM in the past.
Exact diagonalization
methods\cite{Dagotto89,Hirsch89,Jolicoeur90,Deutscher93,Capriotti99}
suffer from exponentially growing size of the Hilbert space,
which is especially a problem in two or more dimensions,
while QMC techniques\cite{Loh90}
suffer from the sign problem for frustrated lattices, and
projected entangled pair states (PEPS) \cite{PEPS,Orus14} for this model
is complex and computationally costly, even though, in principle, PEPS has
good computational scaling properties in 2D.
More recently some numerical methods have been developed that are
especially useful for frustrated systems and applied to the THM.
For example, large-scale parallel tempering Monte
Carlo\cite{Seabra11} and some tensor networks methods including
entangled-plaquette states\cite{Mezzacapo10} and multi-scale entanglement
renormalization ansatz (MERA).\cite{Harada12} 


On the other hand, matrix product states (MPS)\cite{MPS} have been around in various
guises for a long time and are a good representation of the ground-state of
1D chains and few-leg ladders.
MPS exploits the
locality of the interactions for 1D ladders, and computes a truncated Hilbert space
that is well-suited for describing ground-states, as it satisfies the \emph{area law}
for the bipartite entanglement (see [\onlinecite{Orus14}] and references therein).
In particular, the density matrix renormalization
group\cite{White92,McCulloch07,MPS} (DMRG) method for finding the variational ground-state
is mature and highly efficient.
A recent study\cite{Stoudenmire12} of different numerical methods
suggested that two-dimensional DMRG could be ``one of the most
powerful methods'' for studying quantum lattice systems. 

A study using same method as this paper derived the phase diagram of the
$J_1$-$J_2$ Heisenberg model on a kagome lattice, which also contains
a rich variety of phases,\cite{Kolley14} including a spin liquid and magnetically ordered
states. The THM on a 3-leg ladder has been previously studied for anisotropic NN 
interactions  with a magnetic field,\cite{Chen13} and a phase diagram
has been obtained. At the isotropic point, corresponding to $\theta=0$ in our
notation, it was shown 
that the introduction of a magnetic field,
$-h \sum_{i}S^z_i$, causes the $120^{\circ}$ 
state to evolve into commensurate planar phases with Y- and V-shape spin ordering
on either side of a $1/3$ magnetization plateau.\cite{Chen13} 

\section{Methods: MPS and DMRG\label{sec:methods}}

In this paper we employ the MPS ansatz,
keeping up to $m=1000$ basis states, using the DMRG method for obtaining the ground-state
wavefunction. The Hamiltonian has $SU(2)$ symmetry, 
\begin{equation} 
[H, \vektor{S}]=0 \; .
\end{equation}
Exploiting this symmetry in the calculations gives a significant
improvement in efficiency, by reducing the dimension of the computational Hilbert space.
Using $m=1000$ SU(2)-symmetric basis states is equivalent to
$m\approx3000$ states with no (or just Abelian $U(1)$) symmetry.
We performed both finite DMRG and infinite
DMRG\cite{McCulloch08} (iDMRG) calculations. The latter exploits
translational symmetry available in the thermodynamic
limit.

Because MPS is fundamentally a 1D ansatz, to apply it to 2D models
a mapping is necessary. We map the 1D chain of spins into a size
$N=L \times 3$ chain in the YC configuration as shown in \fref{fig:lattice}, where $L$ is
the length of the 3-leg cylinder. The
computational cost will scale approximately linearly with length, but still
exponential with width, which is a limitation of this
method. The model is on a cylinder, i.e. we use open
boundary conditions (OBC) 
in the long (horizontal) direction and periodic
boundary conditions (PBC) in the short (vertical) direction,
so that the total number of NN or NNN bonds in the lattice is
$3N-12$. From finite
size scalings of the energy and other order parameters, we found that
sizes up to $30 \times 3$ 
are large enough to scale finite
results properly into the thermodynamic limit.

In the case of iDMRG, one can classify all possible wrappings of the
triangular lattice on an infinite 3-leg cylinder, using a standard
notation developed for single-wall carbon nanotubes.\cite{Wildoer98}
We use the wrapping vector $\vektor{C}_0=(-3,3)$ in this notation.
The unit vectors, $\vektor{a_1}$ and $\vektor{a_2}$, used to specify $\vektor{C}_0$,
are shown in \fref{fig:lattice}. $\vektor{C}_0$ preserves the tripartite
symmetry on the infinite lattice. The pitch angle of this wrapping
method is $\phi_0 = 90^{\circ}$. The matrix product operator\cite{McCulloch07} (MPO) 
representation of the Hamiltonian has a 3-site unit-cell 
in the direction of $\vektor{C}_0$. One can show that $\vektor{C}_0$
is the shortest possible wrapping vector that preserves tripartite symmetry.
The minimum unit-cell of the wavefunction however is 18 sites, as the smallest
even size that preserves tripartite symmetry.

\subsection{Error Analysis}

\begin{figure}
  \begin{center}
    \includegraphics[width=\columnwidth]{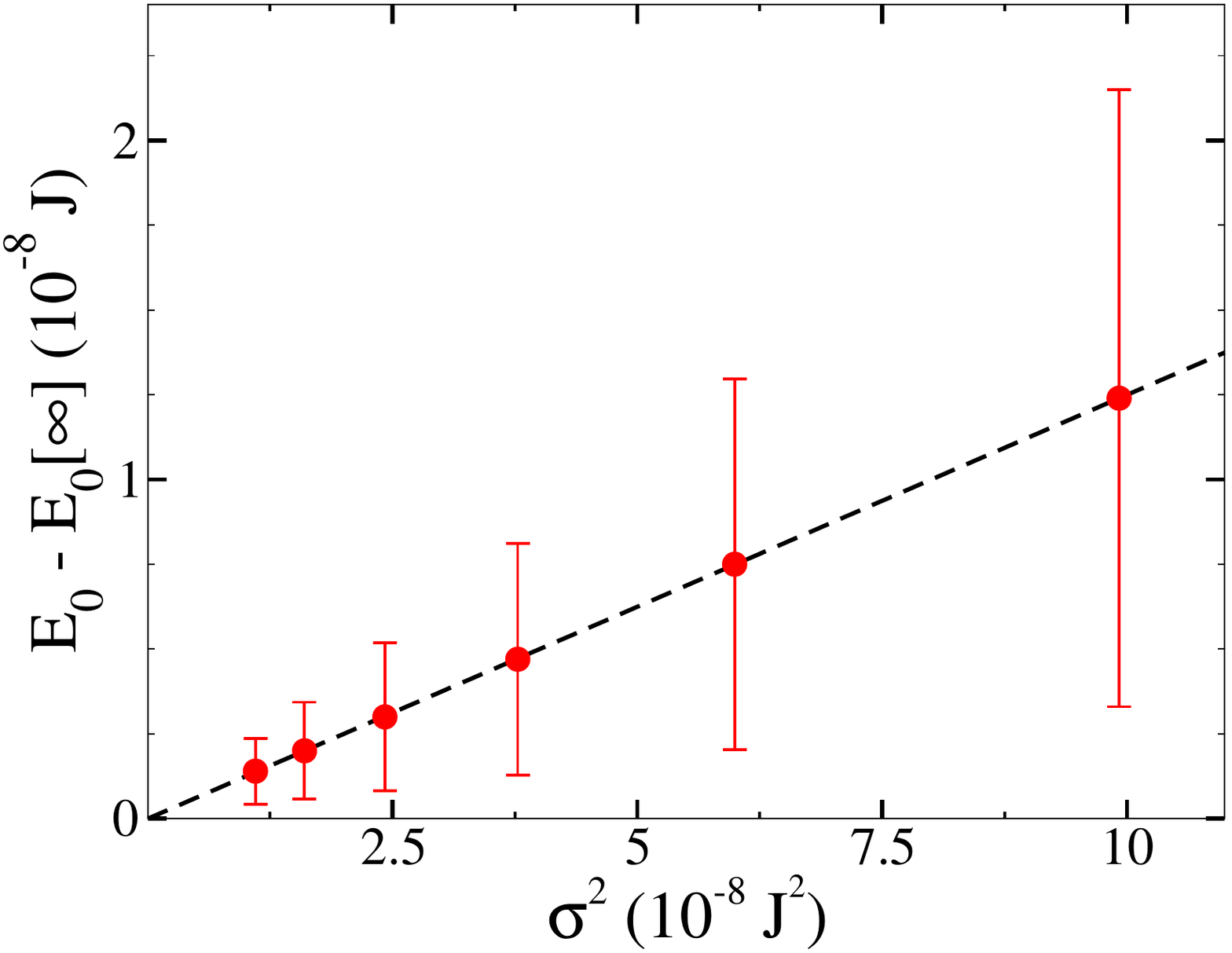}
    \caption{(Color online) Energy versus variance for a $30\!\times\!3$ 
      cylinder with $\theta=25^\circ$.
      The linear extrapolation gives a good approximation for the exact ground-state energy.
      \label{fig:EnergyVsVariance}}
  \end{center}
\end{figure}

We use the variance to calculate systematic errors in the DMRG results.
E.g.~in the case of energy, we have $\sigma^2_E = \langle \psi_v | (H-E)^2
 | \psi_v \rangle$. For energy errors, one needs
to plot energy versus variance step-by-step for different numbers of states, $m$.
The behavior of $E$ versus $\sigma^2_E$ is expected to be linear. Any significant
deviation from this linearity indicates that the DMRG calculation has not converged,
possibly due to an insufficient number of basis states. An example of this calculation
is shown in \fref{fig:EnergyVsVariance}, for a $30 \times 3$ lattice with $\theta=25^\circ$,
for $m$ between 500 and 1000. The ground-state energy extrapolated 
to the $m \rightarrow \infty$ limit
is $E_0[\infty] = -46.94877331266(2)$.
This method is similar to, but more robust than, the energy versus truncation error
scaling that is typically used in DMRG calculations.\cite{McCulloch07}

Throughout this paper the results are all converged with relative errors
$\sim 10^{-12} - 10^{-8}$. Errors are smaller than
symbol size for all plots except for the finite-size extrapolation of the spin gap 
in \fref{fig:SpinGap_Binder}(a), where we show explicit error bars.

\section{Phase Diagram}
\label{sec:PhaseDiagram}


\begin{figure}
  \begin{center}
    \includegraphics[width=\columnwidth]{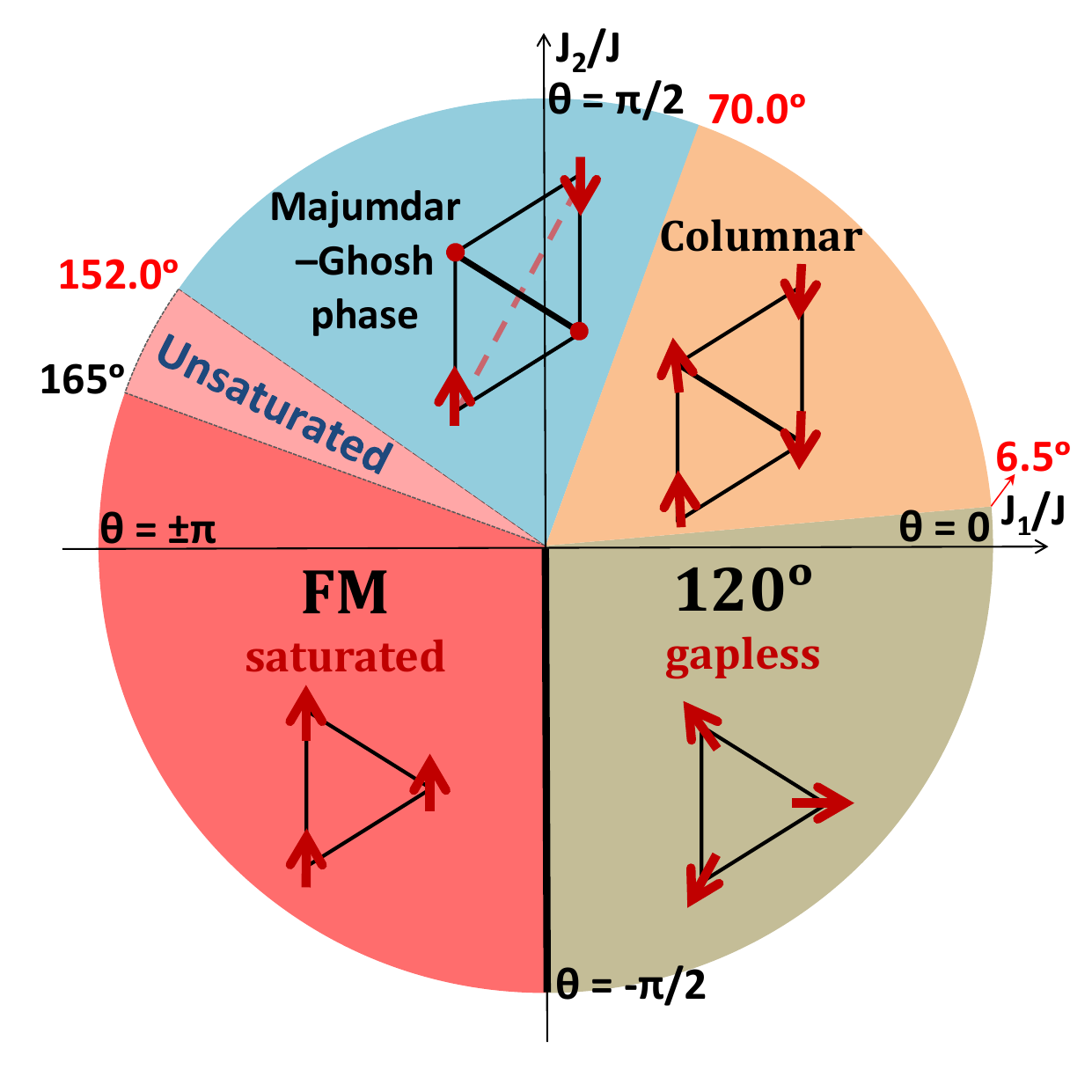}
    \caption{(Color online) The calculated phase diagram of the $J_1$-$J_2$ THM
      on a 3-leg cylinder. The phase transitions are indicated to a
      resolution of $0.5^\circ$. All transitions are second-order except for
      $\theta = -{\pi}/{2}$, which is first-order (marked by a
      thick black line).\label{fig:PhaseDiagram}}
  \end{center}
\end{figure}

The calculated phase diagram of the $J_1$-$J_2$ THM is shown in
\fref{fig:PhaseDiagram}. The dominant short-range ordering is sketched in the form of
triangular or rhombic plaquettes. 
The model contains four well-defined phases. The different
phases were determined by studying the ground-state energy
(\sref{sec:energy}), spin-spin correlation functions (\sref{sec:corr}), 
the chirality (\sref{sec:chirality}), $120^{\circ}$ order parameter (\sref{sec:120}), 
spin gap, dimer order parameter, and Binder cumulant (\sref{sec:critical}). 

\begin{figure}
  \begin{center}
    \includegraphics[width=\columnwidth]{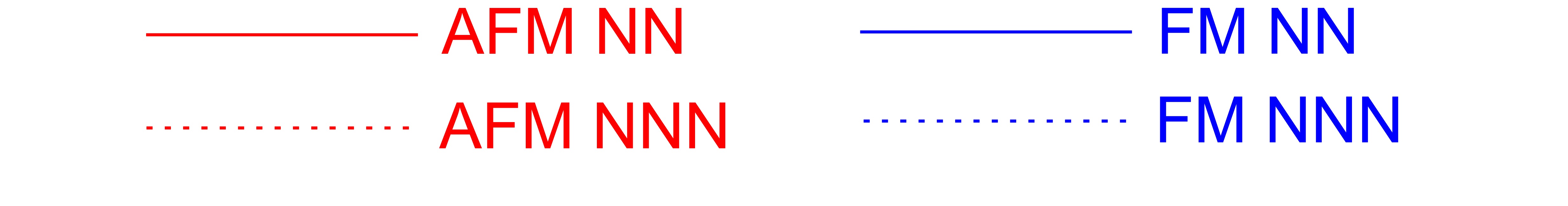}\\
    \includegraphics[width=0.48\columnwidth]{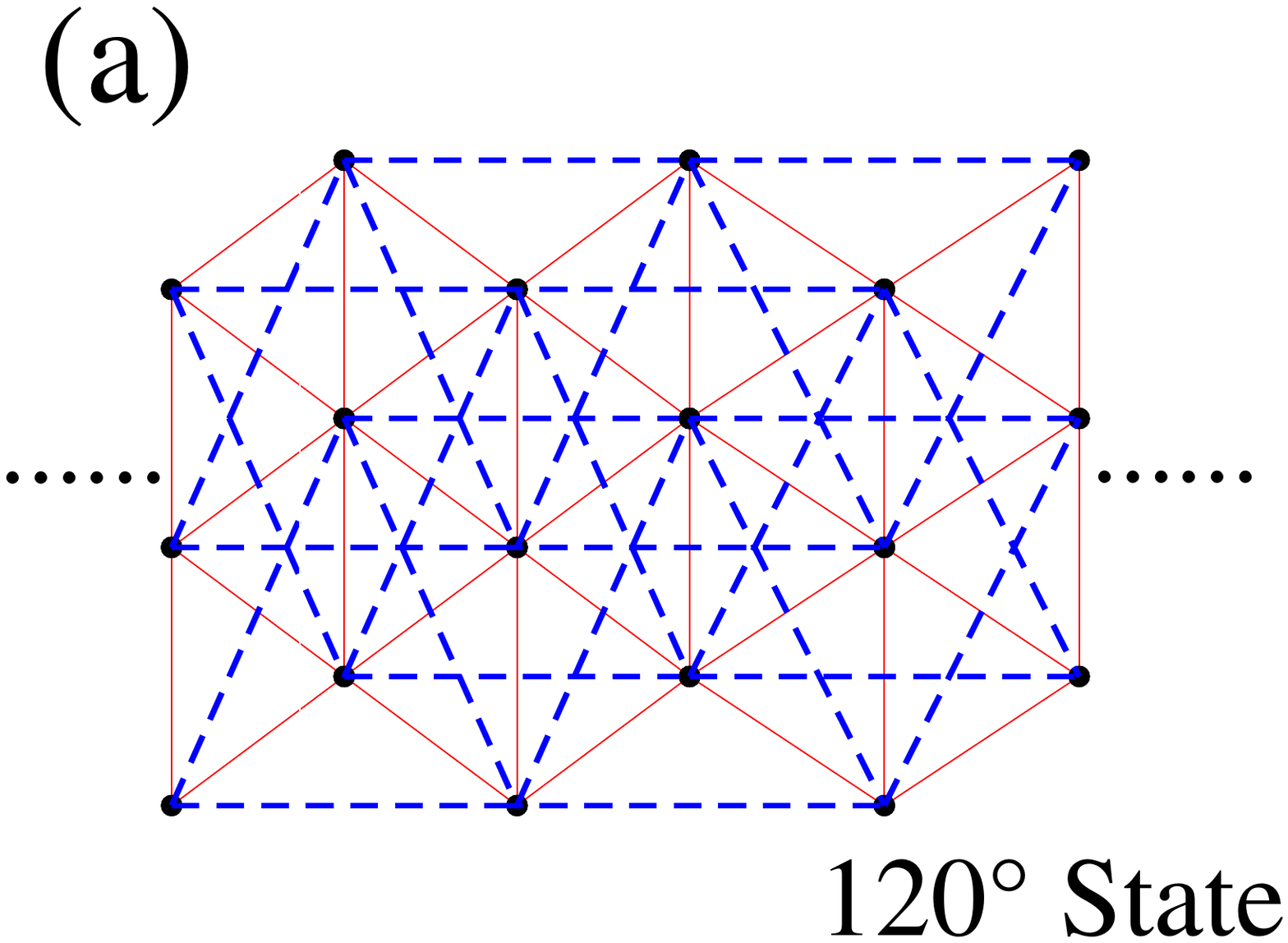}
    \includegraphics[width=0.48\columnwidth]{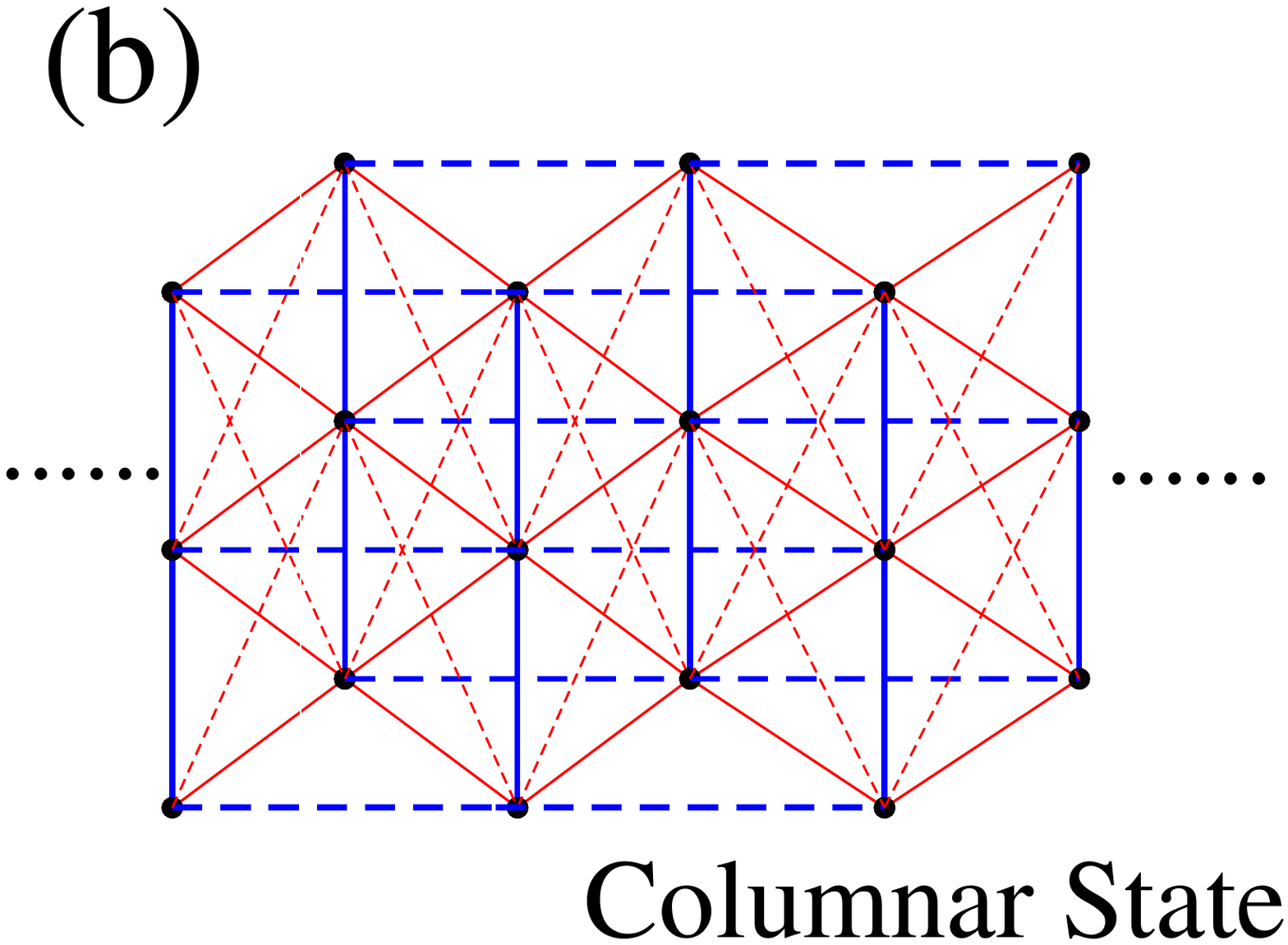}\\
    \includegraphics[width=0.48\columnwidth]{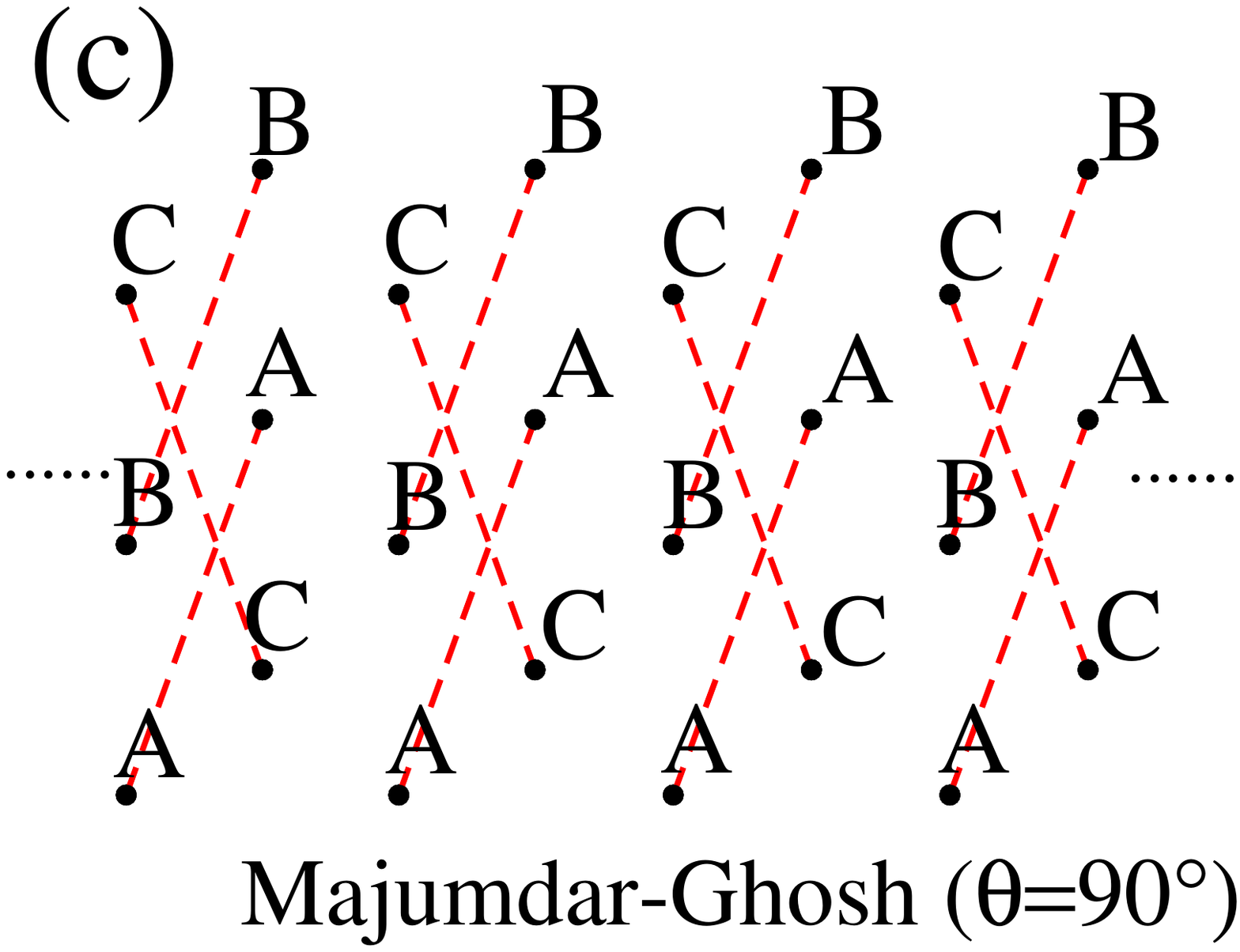}
    \includegraphics[width=0.48\columnwidth]{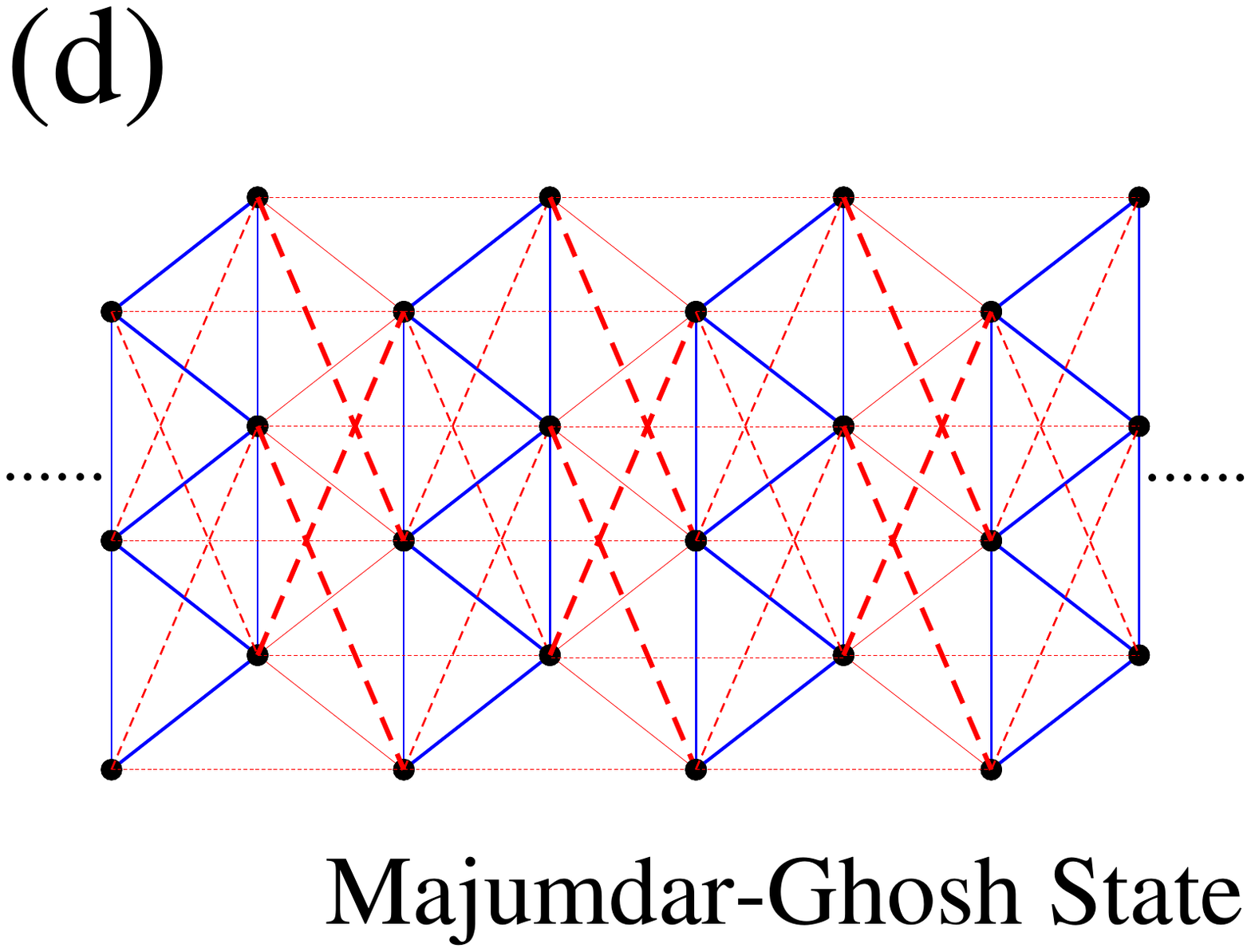}
    \caption{(Color online) Visualizations representative of different phases of the THM
      on a cylinder. Thicker lines represent stronger bonds
      with red indicating antiferromagnetic bonds, blue indicating ferromagnetic. 
      \label{fig:RLV}}
  \end{center}
\end{figure}

In order to better visualize the nature of the short-range correlations in
each phase, \fref{fig:RLV} shows the NN and NNN bonds,
colored according to the value and sign of the spin-spin
correlation. The four phases are:

\begin{enumerate}

\item A \textbf{$\mathbf{120^{\circ}}$ state} (cf. \fref{fig:RLV}(a)) that
exists in the fourth quadrant of \fref{fig:PhaseDiagram}.
The $120^{\circ}$ state is critical (see \sref{sec:critical} below), 
with power-law correlations and gapless excitations.
This is in contrast to the 2D model, which has long-range
tripartite magnetic ordering. However long-range magnetic ordering
is forbidden in our 3-leg cylinder
due to the Mermin-Wagner theorem\cite{MerminWagner66}, which excludes
$SU(2)$ symmetry-broken long-range order in 1D.
The $120^{\circ}$ state is parity-symmetric (P-symmetric), time-reversal-symmetric 
(T-symmetric), and planar (see \sref{sec:chirality} below). We find that the NN 
spin-spin correlation functions are C$_6$ symmetric in this phase (cf. Fig. 
\ref{fig:correlations}); however chiral correlators are $C_3$ symmetric,
reflecting the antiferrochiral ordering (cf. section \ref{sec:Chiral}).
This state persists in
the first quadrant up to a quantum critical point at
non-trivial $\theta_c \simeq 6.5^{\circ}$. The
existence of $120^{\circ}$ state is consistent with spin-wave results of Jolicouer et
al.\cite{Jolicoeur90} Although the transition point of spin-wave
calculations is located at $\theta_{SWT} = \tan^{-1} \frac{1}{8} \simeq 7.125^{\circ}$
compared to our value of $6.5^{\circ}$.

\item Upon increasing $\theta > 6.5^{\circ}$,
the system changes phase to a two-sublattice commensurate 
spin state with a \textbf{columnar} structure (cf. \fref{fig:RLV}(b)), which
is also gapless.
This is consistent with the 2D model, which has long-range columnar 
order.\cite{CCMSpinLiquid,QMCSpinLiquid1,QMCSpinLiquid2}
However this ordering is forbidden in 1D for the same reason as the $120^\circ$ state.
The columnar state is quasi-long-range, C$_6$
rotational symmetry broken, P-symmetric, T-symmetric, and planar. 
This phase can be thought of as a
planar version of the standard G-type
antiferromagnetism.\cite{Getzlaff07}

\item At $\theta \simeq 70.0^\circ$, there is a phase transition
to a \textbf{NNN Majumdar-Ghosh state}. 
In this phase the system forms strong AFM
bonds (dimers) along NNN bonds. Because of the finite width of the 
ladder and the periodic boundary conditions in the short direction,
each site is NNN to some other site twice (e.g., the exchange interaction
between sites 38 and 42 is twice that between site 38 and 44, cf. 
\fref{fig:Lattice}. 
At $\theta=\pi/2$, the the model is composed of three uncoupled
sublattices in the form of 2-leg spin ladders. 
The double counting of the NNN bonds means that
the Majumdar-Ghosh Hamiltonian is realised in each sublattice, 
leading to three copies of the two-fold degenerate Majumdar-Ghosh state
with long-range dimer order, shown in \fref{fig:RLV}(c)).
The Majumdar-Ghosh state is robust
to small perturbations when one turns on the $J_1$ interactions,
and evolves into the general form shown in
\fref{fig:RLV}(d), with weak NN bonds, either antiferromagnetic or ferromagnetic
corresponding to the sign of $J_1$. 
We find, numerically, that this 
 state persists throughout
a large region in the first and second quadrants of  \fref{fig:PhaseDiagram}.
The Majumdar-Ghosh state has short-ranged correlations (cf.~\fref{fig:CorrLength_iDMRG}), 
and is C$_6$ rotational symmetry broken, translational symmetry broken,
P-symmetric, T-symmetric, and planar. 

\item Upon further increasing of $\theta$, the system undergoes a second-order phase
transition at $\theta_c = 152.0^{\circ}$ (see \sref{sec:critical} below).
In a narrow region,
$152^{\circ} < \theta < 165^{\circ}$ of \fref{fig:PhaseDiagram}, the ground-state
is a partially polarized \textbf{ferromagnet} that saturates to
complete ferromagnetism for $\theta > 165^\circ$.

\end{enumerate}


\subsection{Limiting Cases}

In our parameterization of the Hamiltonian, $\theta = 0$ is equivalent to $J_2=0$, 
and is simply the nearest-neighbor model. The ground-state is the
$120^{\circ}$ state, in agreement with the semi-classical 
approach,\cite{Jolicoeur90} with wave vector $\vektor{Q} = (2\pi/\sqrt{3},
2\pi/3)$ in our notation. 

For $\theta = 90^\circ$ ($J_1 = 0$),
the model has only NNN interactions.
This state is composed of three uncoupled spin
ladders, one in each tripartite sublattice, forming a perfect Majumdar-Ghosh 
state of alternating singlet dimers.\cite{MajumdarGhosh,AuerbachBook} 
The formation of this phase is a direct consequence of the 3-leg form
of the lattice, \fref{fig:lattice}, which is wrapped around a cylinder
resulting in 3 independent zig-zag spin chains with NN coupling $J_2$, and the
double-counted bonds around the periodic boundary give an NNN
coupling of $2 J_2$. 
Thus, this state appears because of the restricted geometry of the 3-leg ladder. 
On the other hand, in the 2D limit the Hamiltonian is
instead three copies of the $\theta=0$ model, hence the ground-state will contain
three copies of the $120^\circ$ state, one on each sublattice, 
and a small $J_1$ will couple the otherwise independent sublattices.
Thus the small-$J_1$ behavior for few-leg ladders is rather different to the bulk
2D behavior.

For $0 < \theta < \tan^{-1} (\frac{1}{8})$ the 2D model at the classical level 
($S \rightarrow \infty$) has a $120^{\circ}$ ground-state\cite{Chubukov92} and for
$\theta > \tan^{-1} (\frac{1}{8})$ it has a 4-sublattice AFM N\'{e}el phase with an infinite
manifold of degenerate ground-states, selected by the ``order from disorder'' mechanism.
Quantum fluctuations break this
degeneracy, and the quantum model has a two-sublattice columnar (collinear) 
N\'eel state.\cite{CCMSpinLiquid,QMCSpinLiquid1,QMCSpinLiquid2} It is worth mentioning that the 
selection of the collinear order from the 4-sublattice classical order, can be understood 
analytically using group-symmetry analysis.\cite{Lecheminant95}

It is straightforward to show that for the classical $J_1$-$J_2$ THM, if one enforces the 
tripartite symmetry everywhere using a repeated
3-site unit-cell, the ground-state
phase is simply ferromagnetic (FM) for $J_1 < 0$ and
the $120^{\circ}$ state for $J_1 > 0$, independent of $J_2$.

\subsection{Ground-State Energy\label{sec:energy}}

\begin{figure}
  \begin{center}
    \includegraphics[width=\columnwidth]{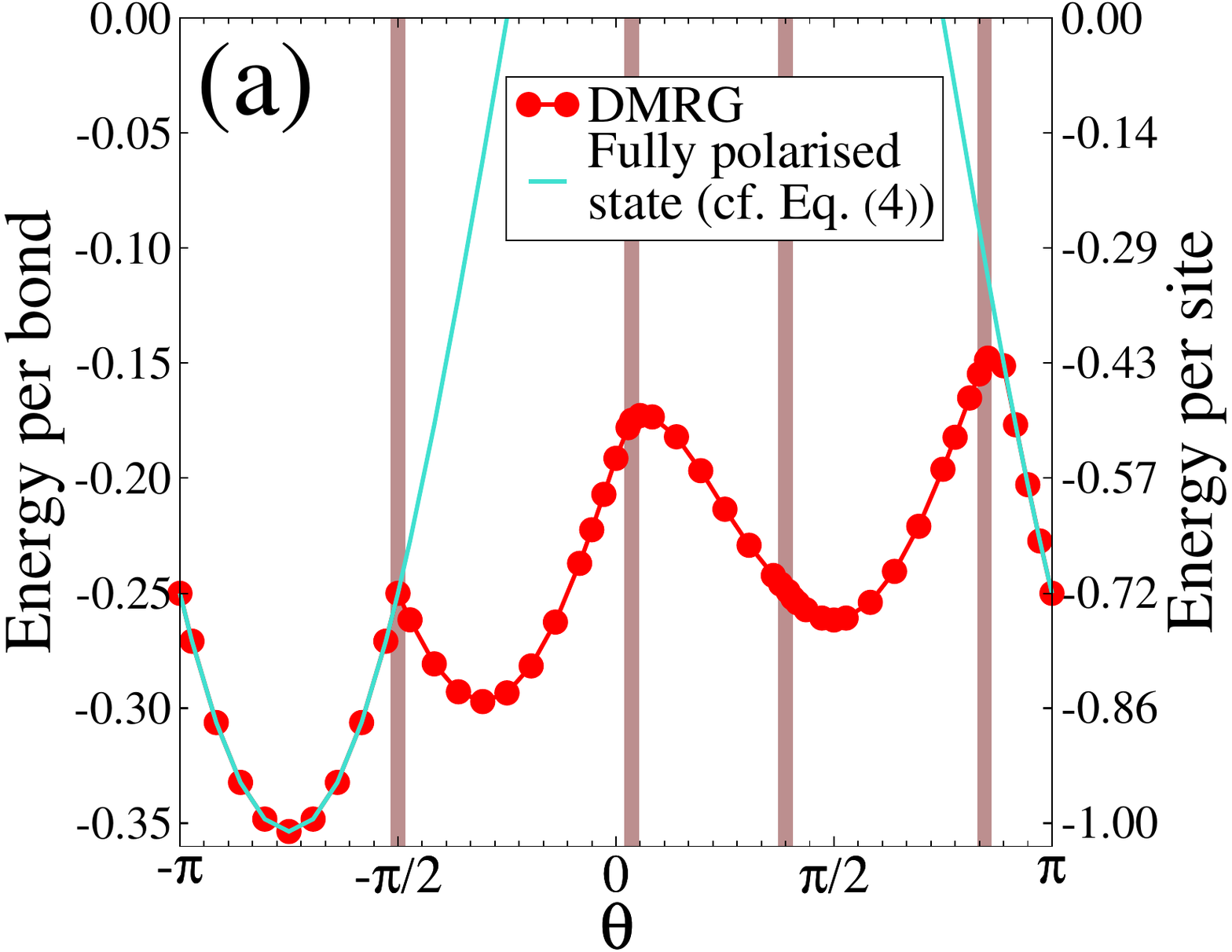}\\
    \includegraphics[width=\columnwidth]{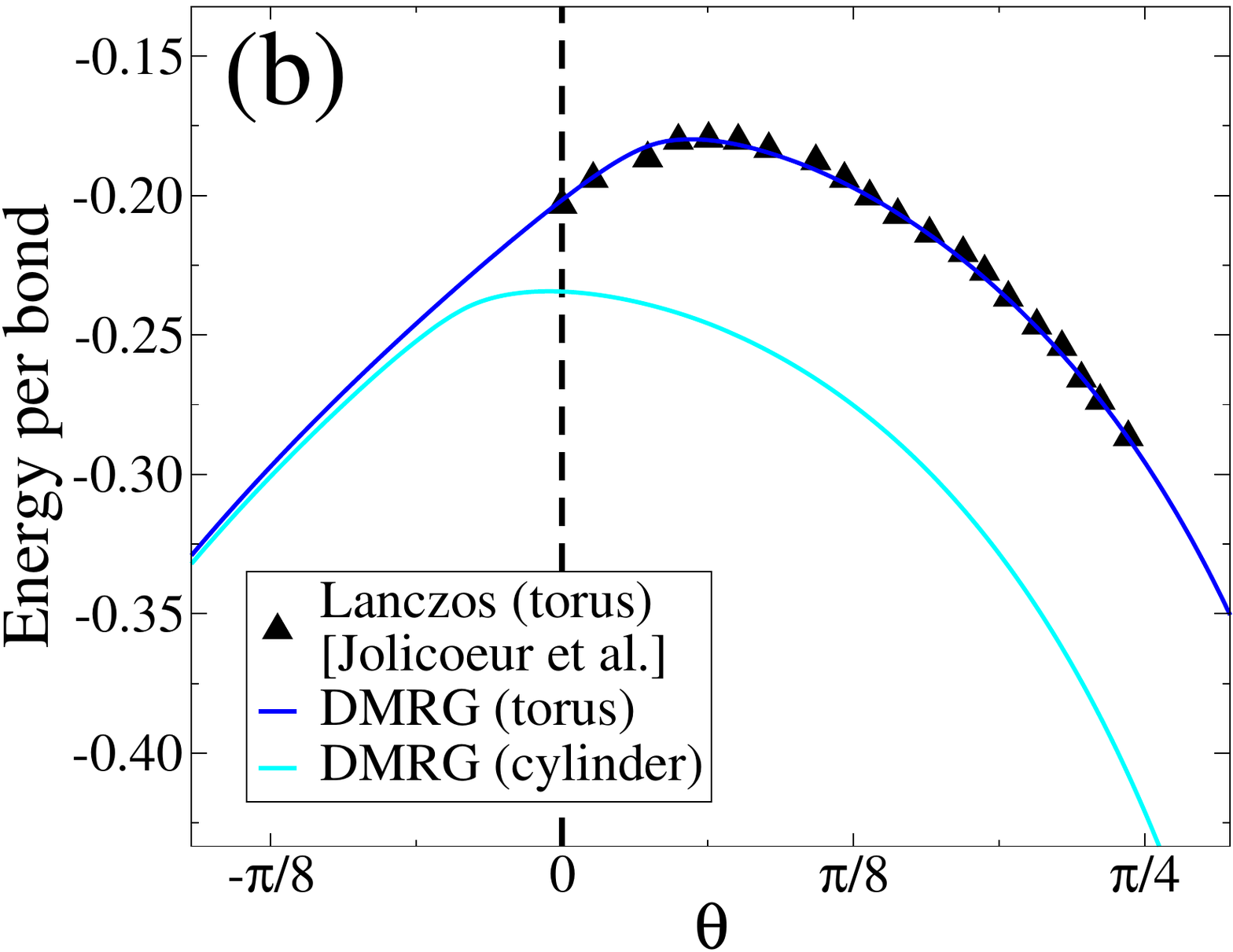}
    \caption{(Color online) (a) Ground-state energy of the THM on a
      $30\!\times\!3$ cylinder. The transition at
      $\theta=-{\pi}/{2}$ is sharp and consistent with a first-order
      phase transition (cf. \sref{sec:critical}). The transition near
      $\theta\approx160^{\circ}$ suggests a second order phase
      transition (cf. \sref{sec:critical}). Brown stripes indicate
      phase transitions. (b) Comparison of the energies of Lanczos and DMRG
      for the THM on a 4$\times$3 lattice. Lanczos results are
      from Jolicoeur et al.\cite{Jolicoeur90} DMRG and Lanczos show
      excellent agreement, but the boundary conditions clearly have a
      significant effect on the energy of this small
      lattice.\label{fig:energy}}
  \end{center}
\end{figure}

In this section, we benchmark our results for the ground-state energy
per nearest-neighbor ($J_1$) bond, $\ve_0$. This is shown in \fref{fig:energy}.
The energy per NN bond in the fully polarised ferromagnet is,
\begin{equation}
  \ve_{FM} = \frac{1}{4} ( \sin\theta + \cos\theta ) \; ,
\end{equation}
which is shown in turquoise in \fref{fig:energy}(a).

There is a sharp transition appearing at $\theta=-\pi/2$,
coinciding with the change from FM to $120^{\circ}$ state. The cusp
suggests a first-order phase transition, which is confirmed by the local
magnetization and order parameters. This is the only first-order transition
that we find in the model, and is indicated by the thick black
line in \fref{fig:PhaseDiagram}. 
On the right-hand side of \fref{fig:energy}(a), at
$\theta_c \simeq 152.0^\circ$, the derivative is continuous
indicating that the transition from the Majumdar-Ghosh state to the FM is
second-order, which we verified by calculating the magnetization (see \sref{sec:critical} below).

\fref{fig:energy}(b) is a comparison of DMRG energies with Lanczos
results of Jolicoeur et al.\cite{Jolicoeur90} They simulated the same
model on a 12-site lattice with PBC in both directions, which is
equivalent to a $4\times3$ torus in our representation. The choice of
wrapping vector around the torus has little effect
as long as the lattice translational and tripartite symmetries
are preserved. Our DMRG results are in very good agreement with
these Lanczos results.

\begin{table*}
  \caption{Comparison of the ground-state energies from different methods for the THM at $\theta=0$. 
  \label{table:EnergyComparison}}
  \begin{center}
    \begin{ruledtabular}
      \begin{tabular}{ cccc }
        Method & Cluster size & Boundary conditions & Energy per bond, $\ve^{\theta=0}_0$ \\ \hline
        \small DMRG (this work) \normalsize & 4$\times$3 & torus & $\leq -0.20164623520324(1)$ \\ 
        \small DMRG (this work) \normalsize & 60$\times$3 & cylinder & $\leq -0.19053054(3)$ \\ 
        \small DMRG (this work) \normalsize & \specialcell[c]{extrapolated to thermodynamic limit\\($L\times3$ lattices with $L \leq 60$)} & cylinders & $-0.189(2)$ \\
        \small iDMRG (this work) \normalsize & infinite & 3-leg cylinder & $\leq -0.189715084187(2)$ \\ \hline
        \small Schwinger boson\cite{Gazza93} \normalsize& $N=12$ & torus & $-0.1899$ \\
        \small Ising expansion\cite{Weihong99} \normalsize & theoretically thermodynamic limit & - & $-0.187$ \\
        \small Entangled plaquette states\cite{Mezzacapo10} \normalsize & \specialcell[c]{extrapolated to thermodynamic limit\\(clusters up to size $N=324$)} & torus & $-0.18473(4)$ \\
        \small Coupled cluster method\cite{Li15} \normalsize & extrapolated to thermodynamic limit & - & $-0.1840(1)$ \\
        \small Numerical diagonalization\cite{Nishimori88} \normalsize & \specialcell[c]{extrapolated to thermodynamic limit\\(clusters up to size $N=27$)} & torus & $-0.183 \pm 0.003$ \\
        \small QMC\cite{Capriotti99} & \small extrapolated to thermodynamic limit \normalsize& - & $-0.182(3)$ \\
        \small SWT\cite{Nishimori85} \normalsize & theoretically thermodynamic limit & - & $-0.182$ \\
        \small MERA\cite{Harada12} \normalsize & \specialcell[c]{extrapolated to thermodynamic limit\\(clusters up to size $N=114$)} & torus & $-0.18029$ \\     
      \end{tabular}
    \end{ruledtabular}
  \end{center}
\end{table*}

\tref{table:EnergyComparison} is a comparison between our DMRG energy
and results from previous calculations for $\theta=0$, 
i.e. the NN model in
the $120^{\circ}$ phase. For this point we performed a
larger size calculation on a $60\times3$ cylinder, as there is
no NNN frustration and the DMRG is easier to converge.
The results in \tref{table:EnergyComparison} suggest that the THM on a
cylinder is a good approximation for the full 2D model.

\subsection{Local magnetization}

\begin{figure}
  \begin{center}
    \includegraphics[width=\columnwidth]{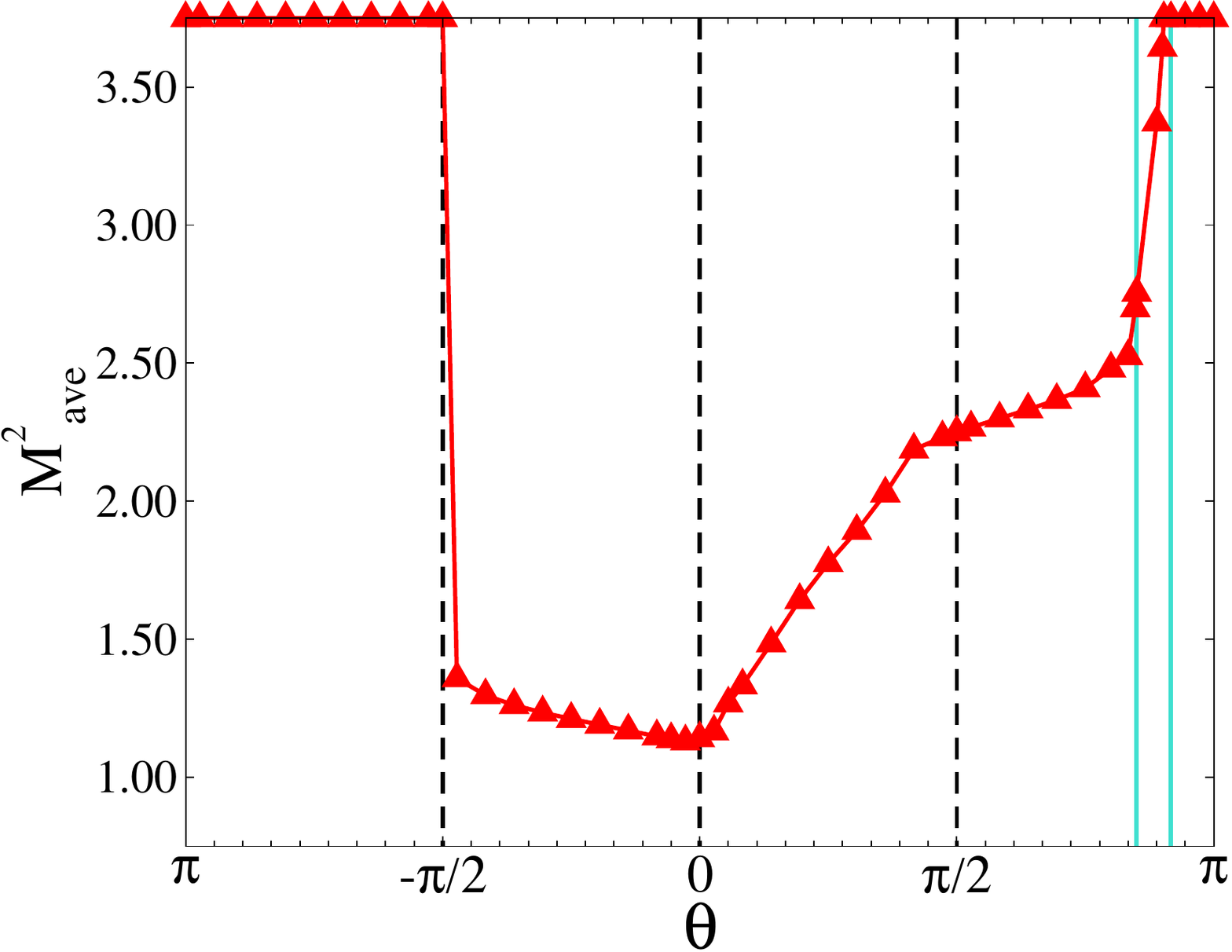}
    \caption{(Color online) Average of squared magnetization per
      plaquette, \eref{eq:mag}, on a 30$\times$3 cylinder.
      The vertical turquoise lines denotes the partially-polarized region
      at the second-order transition into the ferromagnetic state.
      \label{fig:mag_ave}}
  \end{center}
\end{figure}

The squared magnetization per plaquette, $M^2_{ave}$ 
is presented in \fref{fig:mag_ave}. This is calculated from
the square of the local magnetization on a single plaquette,
\begin{equation}
  M^2_{ave} = \frac{1}{N_p}
\sum_{\{A,B,C\}} ( {\bf S}_A + {\bf S}_B + {\bf S}_C )^2 \; ,
  \label{eq:mag}
\end{equation}
where the sum is over all $N_P$ plaquettes with vertices $A$, $B$, and $C$ from
their respective sublattice.
The turquoise lines in \fref{fig:mag_ave} 
indicate the region where we find a partially-polarized
ferromagnetic ground-state. The rapid but smooth change in local
magnetization in this region is consistent with a second-order phase transition.

\section{Spin-spin Correlations}
\label{sec:corr}

In this section, we examine the spin-spin correlation functions. Both
the short-range and long-range behavior gives detailed information on the
phases and phase boundaries. Since there is no long-range magnetism 
(except in the ferromagnetic phase, where the order 
parameter commutes with the Hamiltonian), the correlation function is simply
\begin{equation}
  \label{eq:CorrFunc}
  O_s(i,j) = \langle \vektor{S}_i \cdot \vektor{S}_j \rangle \; ,
\end{equation}
where $i$ and $j$ are the indices specifying spin vertices in
the lattice, \fref{fig:lattice}. 


\subsection{Short-range Correlations}

\begin{figure}
  \begin{center}
    \includegraphics[width=\columnwidth]{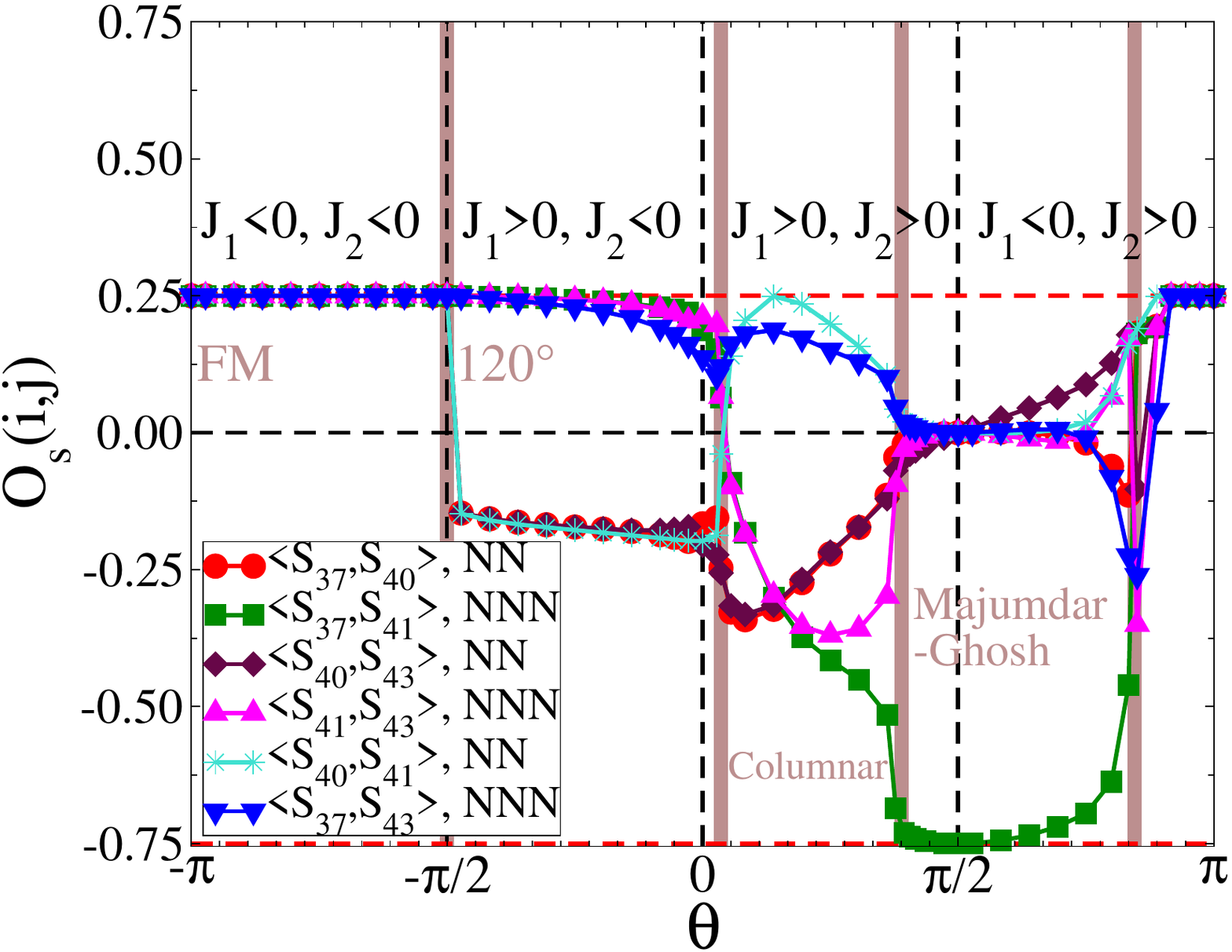}
    \caption{(Color online) Short-ranged spin-spin correlation functions for the
      ground-state on a 30$\times$3 cylinder. $S_n$ represents the
      spin operator for the site $n$ according to the lattice numbering in
      \fref{fig:lattice}. Dashed red lines are the limits of the
      correlation functions for spin-$1/2$ particles, 
      $-3/4 \leq O_s(i,j) \leq 1/4$. Brown stripes indicate phase
      transitions.
      \label{fig:correlations}}
  \end{center}
\end{figure}

To identify the bulk properties of the ground-state, we plot six reference correlation
functions in \fref{fig:correlations}.
There are the short-range correlations calculated for
the central few sites of the $30\times3$ cylinder. The edges
of the lattice show non-negligible boundary effects, however away from the boundary,
the bulk correlations appear to be  representative of the
thermodynamic limit and agree closely with correlators calculated using iDMRG.
In \fref{fig:correlations}, brown stripes indicate the phase transitions that we have identified.

\subsection{Long-range Correlations}

\begin{figure}
  \begin{center}
    \includegraphics[width=\columnwidth]{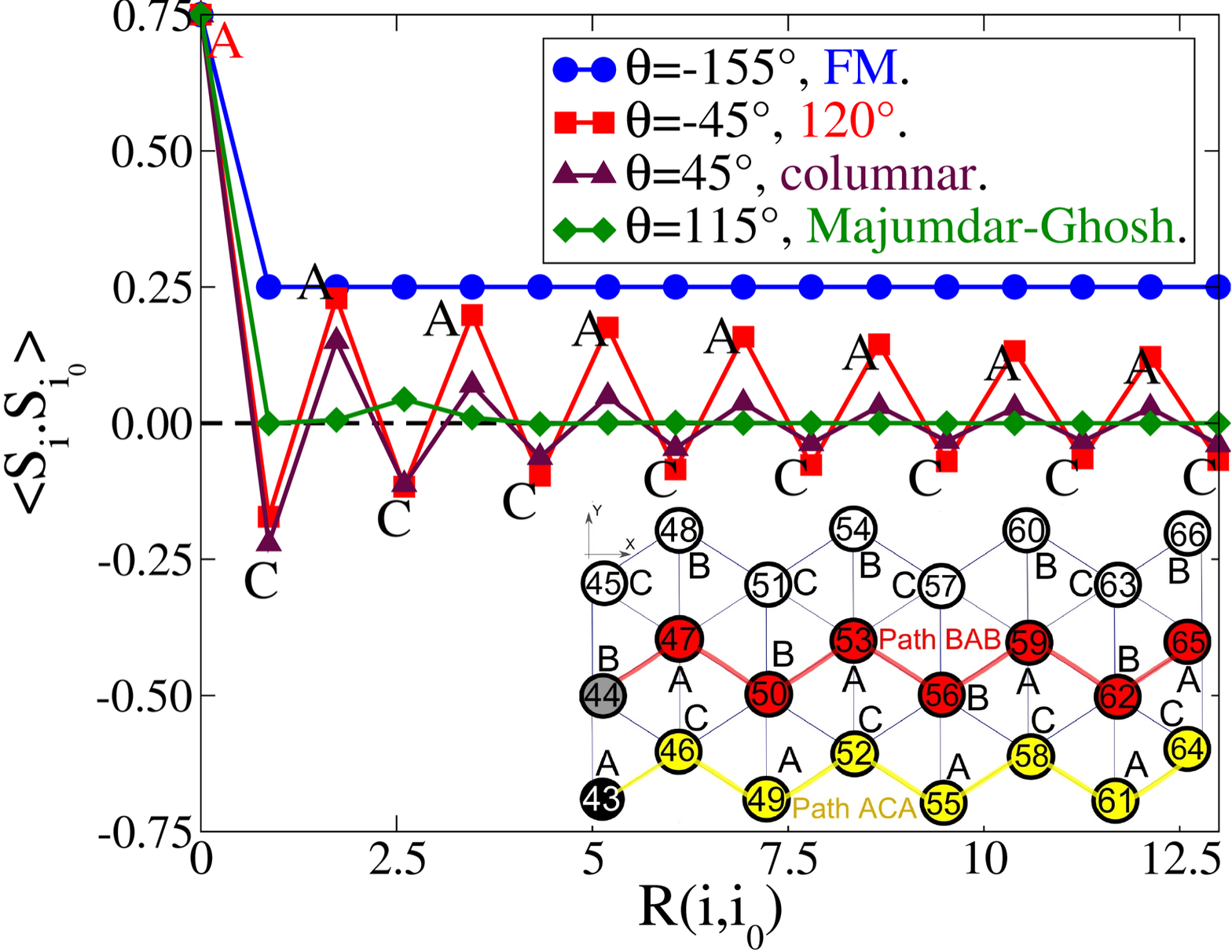}
    \caption{(Color online) Spin-spin correlators for the THM on a
      30$\times$3 cylinder for trial path ACA. $R(i,i_0)$ is spatial
      distance between vertices $i$ and $i_0$ in units of the lattice
      spacing. Two possible paths are shown in the inset.\label{fig:CorrPath}}
  \end{center}
\end{figure}

We now consider the long-range behavior of the spin-spin correlators \eref{eq:CorrFunc}.
One can choose different paths to study distant correlators according to
the lattice geometry, but at long distances the spin-spin correlators are 
independent of the choice of path. The \fref{fig:CorrPath} shows correlators
calculated for the path ACA as shown in the inset. We also
calculated the correlation functions for a number of different
paths. Up to trivial differences caused by the order in which
different sublattices are listed, the results are insensitive to
the path followed. The 
results suggest that the $120^{\circ}$ and columnar states are quasi-long-range and
the Majumdar-Ghosh state contains only short-ranged spin-spin correlations.

\begin{figure}
  \begin{center}
    \includegraphics[width=\columnwidth]{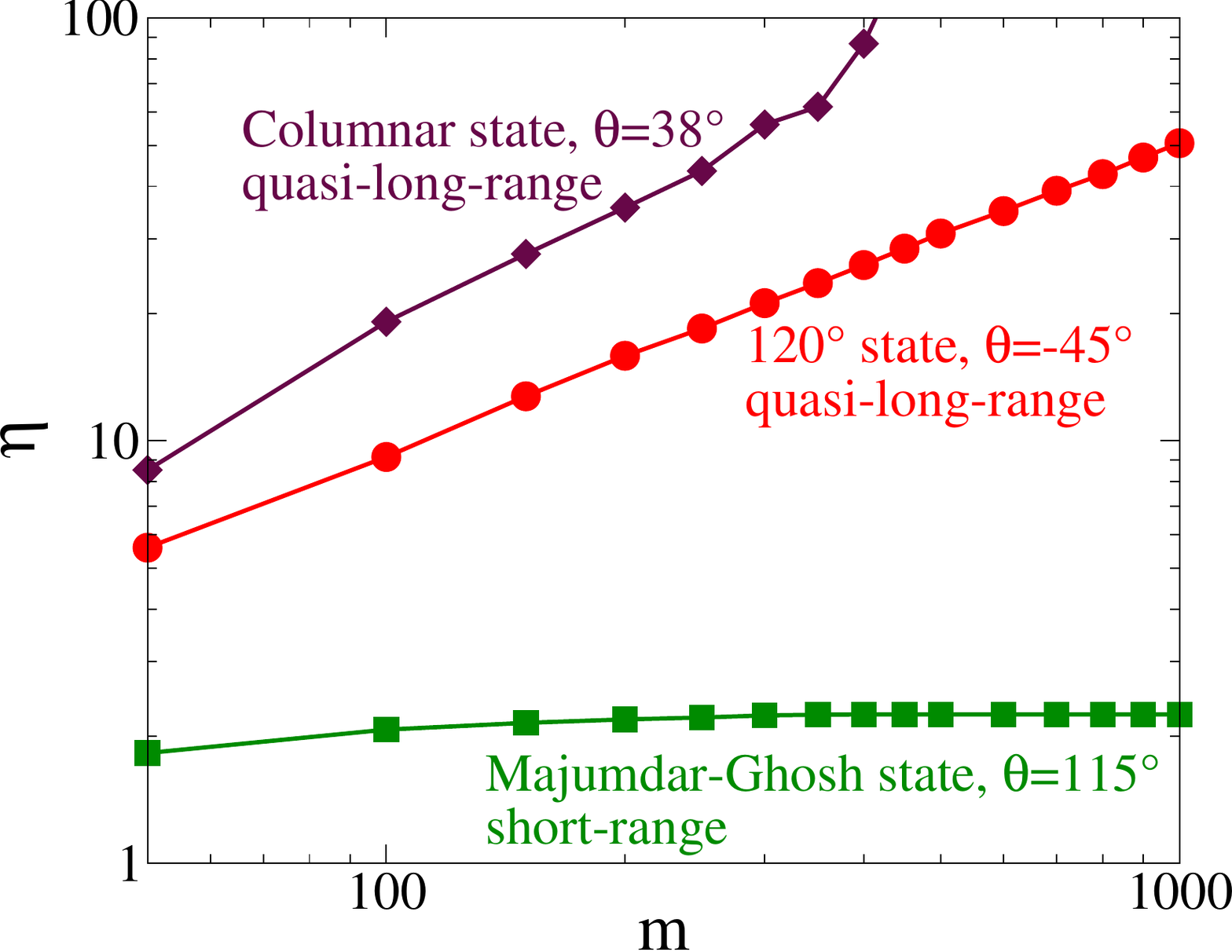}
    \caption{(Color online) iDMRG results for the correlation length. 
      $\eta$ is the correlation length, calculated from the next-leading eigenvalue of the
      transfer matrix, \eref{eq:CorrLength}. 
      Power-law growth $\eta \propto m^\kappa$ indicates gapless quantum critical
      excitations with power-law correlations, while the saturation of the
      correlation length in the Majumdar-Ghosh region indicates that this 
      phase is short-range and gapped.
      \label{fig:CorrLength_iDMRG}}
  \end{center}
\end{figure}

Using iDMRG, one can directly extract the correlation length from
spectrum of the transfer matrix. If $\Lambda$ is the largest magnitude
eigenvalue in the transfer matrix smaller than $1$, then the correlation length
$\eta$ is obtained from
\begin{equation}
  \label{eq:CorrLength}
  |\Lambda| = e^{a_0 / \eta} \; ,
\end{equation}
where $a_0$ is the size of the iDMRG unit-cell. Upon increasing
the number of states, $m$, the observation of power-law growth of the
correlation length indicates a gapless phase, whereas saturation
of $\eta$ is a sign of a gapped phase.\cite{Luca08,Vid15} The result for the correlation
length of the $120^\circ$, columnar, and Majumdar-Ghosh phases are shown in
\fref{fig:CorrLength_iDMRG}. 
This is consistent with finite DMRG results of \fref{fig:CorrPath}, where
the $120^{\circ}$ and columnar states are quasi-long-range and the Majumdar-Ghosh
phase has a finite correlation length.

\section{Chirality\label{sec:chirality}}

For several decades, there has been much 
discussion\cite{GroupChiral,Dagotto89,Jolicoeur90,Chubukov92,Deutscher93}
on the possibility of chiral order in the 2D model. 
A proper chiral order parameter will detect breaking of
P and T-symmetry of the wavefunction while the
system preserves PT-symmetry. This can be done by looking at
order parameters or correlation functions that are not symmetric
under P or T. 

We studied the chirality using two chiral order parameters
introduced below, \eref{eq:CcCorr} and \eref{eq:CtCorr}, which we
evaluated using finite DMRG. 
The results are presented in \fref{fig:CrossCorr} and
\fref{fig:TripleCorr}. These results show that there is \emph{no} long-range chiral order.
We also directly measured the
parity and time-reversal symmetry of infinite length 3-leg cylinders, 
using infinite DMRG. The procedure for this is to
calculate the overlap per unit cell of the iDMRG wavefunction with its conjugate
or parity-reflected version. Since iDMRG works directly in the thermodynamic limit,
spontaneous breaking of discrete symmetries can occur, and this is a reliable
way to detect $P$ or $T$ symmetry breaking.\cite{McCulloch08,Meisner09,Pollmann12}
The calculated overlap, $f$, between the
P-transformed, and T-transformed, wavefunctions is of the order of $1-f \approx 10^{-8}$ per unit cell,
showing that neither P nor T-symmetry is broken.

\subsection{Vector Chirality\label{subsec:CcCorr}}

To measure the local chirality, we use the cross product between
vertex pairs in a plaquette, while keeping a fixed cyclic order of operators,
\begin{equation} {\bf C}_c[A,B,C] = {\bf S}_A \times {\bf S}_B + {\bf
    S}_B \times {\bf S}_C + {\bf S}_C \times {\bf S}_A \;,
\end{equation}
where $[A,B,C]$ stands for a triangular plaquette
composed of vertices from sublattice A, B, and C.
Note however that the magnitude of the local chirality is \emph{not} a good order parameter, since it
is easy to show that for any spin-$1/2$ system we have,
\begin{equation}
  ({\bf S}_i \times {\bf S}_j + {\bf S}_j \times {\bf S}_k + {\bf S}_k \times {\bf S}_i)^2 =
  -\frac{3}{4} M^2_{i,j,k} + \frac{45}{16} \; ,
\end{equation}
where $M^2_{i,j,k}$ is the square of the local magnetization,
\begin{equation}
  M^2_{i,j,k} = ( {\bf S}_i + {\bf S}_j + {\bf S}_k )^2 \; .
\end{equation}
Hence the magnitude of the cross product is directly related to the local magnetization 
and has no connection to the chirality.

\begin{figure}
  \begin{center}
    \includegraphics[width=\columnwidth]{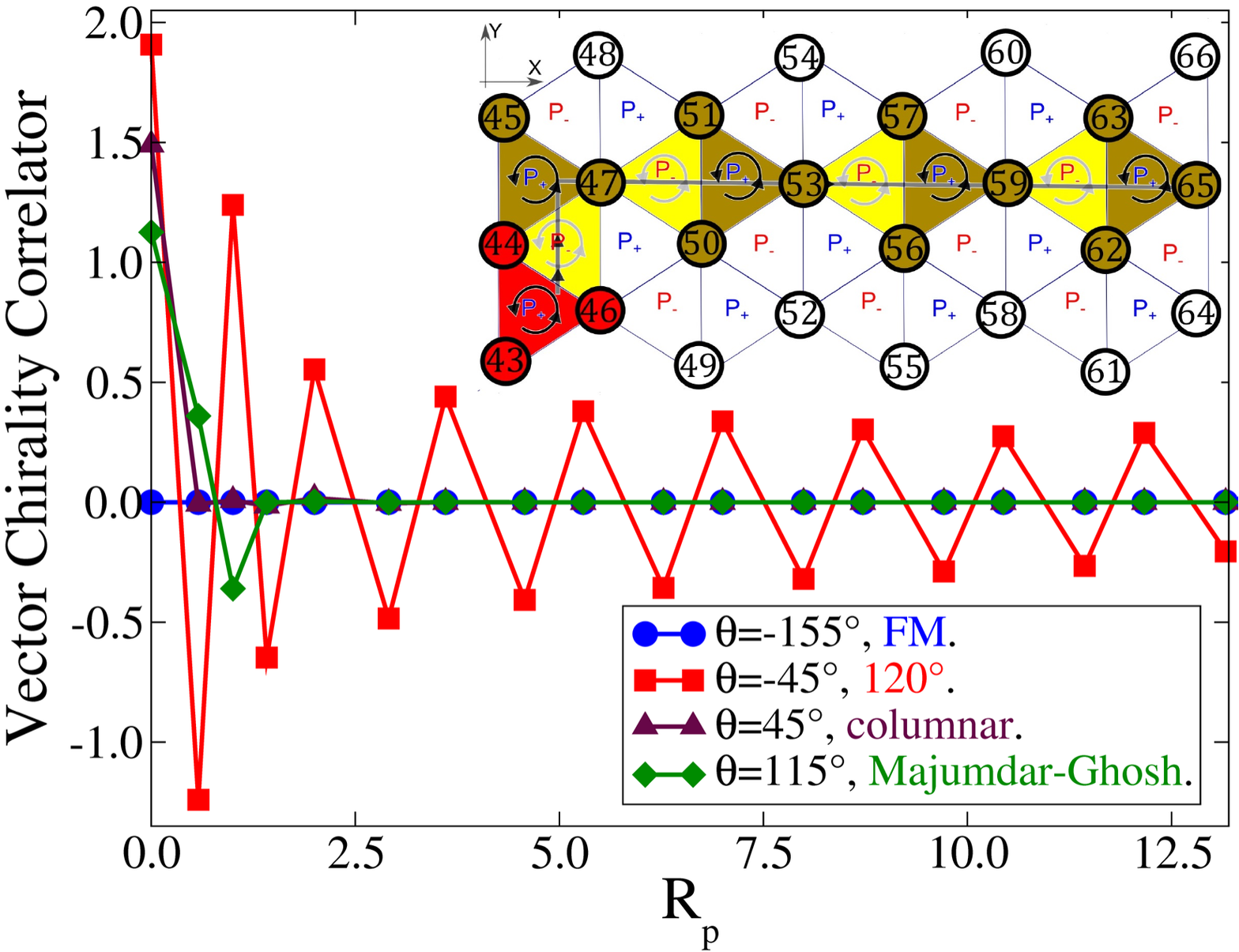}
    \caption{(Color online) Vector chirality correlator, $O_c$,
      \eref{eq:CcCorr}, on a 30$\times$3 cylinder. $R_P$ is
      the distance between the centers of the plaquettes in units of the
      lattice spacing. The ``antiferrochiral'' pattern in the
      $120^{\circ}$ state can be explained by the tripartite symmetry of
      the lattice. Antiferrochirality is clearly broken in the Majumdar-Ghosh 
      state. 
      (inset) The path for
      which the vector chirality correlators were calculated.
      The $P_+ / P_-$ labels indicate the antiferrochiral ordering.
      \label{fig:CrossCorr}}
  \end{center}
\end{figure}

The correlation function of the vector chirality, $O_c$, detects long-range chiral order,
\begin{equation}
O_c(i,j,k;i',j',k') = \langle {\bf C}_c[i,j,k] \cdot {\bf C}_c[i^\prime,j^\prime,k^\prime] \rangle \; .
\label{eq:CcCorr}
\end{equation}
DMRG results for this correlation function are shown in \fref{fig:CrossCorr}. To
calculate these correlators between desired plaqeuttes, we chose a
path that has the maximum number of crossings of plaquette vertices. This path
is shown in the inset of \fref{fig:CrossCorr}. The origin plaquette is
indicated in red. The next two plaquettes respectively have two and one common
vertices with the origin while longer range plaquettes have none. The
results of \fref{fig:CrossCorr} suggest that the $120^{\circ}$ and NNN
Majumdar-Ghosh states are only short-range chiral. There is a
long-range ``antiferrochiral'' pattern in the $120^{\circ}$ state
specified with $P_+ / P_-$ notation in the inset of
\fref{fig:CrossCorr}, which is consistent with the tripartite structure of the lattice.
We calculated the vector chirality for all possible plaquettes, and all
show the antiferrochirality of the $120^{\circ}$ state.

\subsection{Scalar Chirality\label{subsec:CtCorr}}

\begin{figure}
  \begin{center}
    \includegraphics[width=\columnwidth]{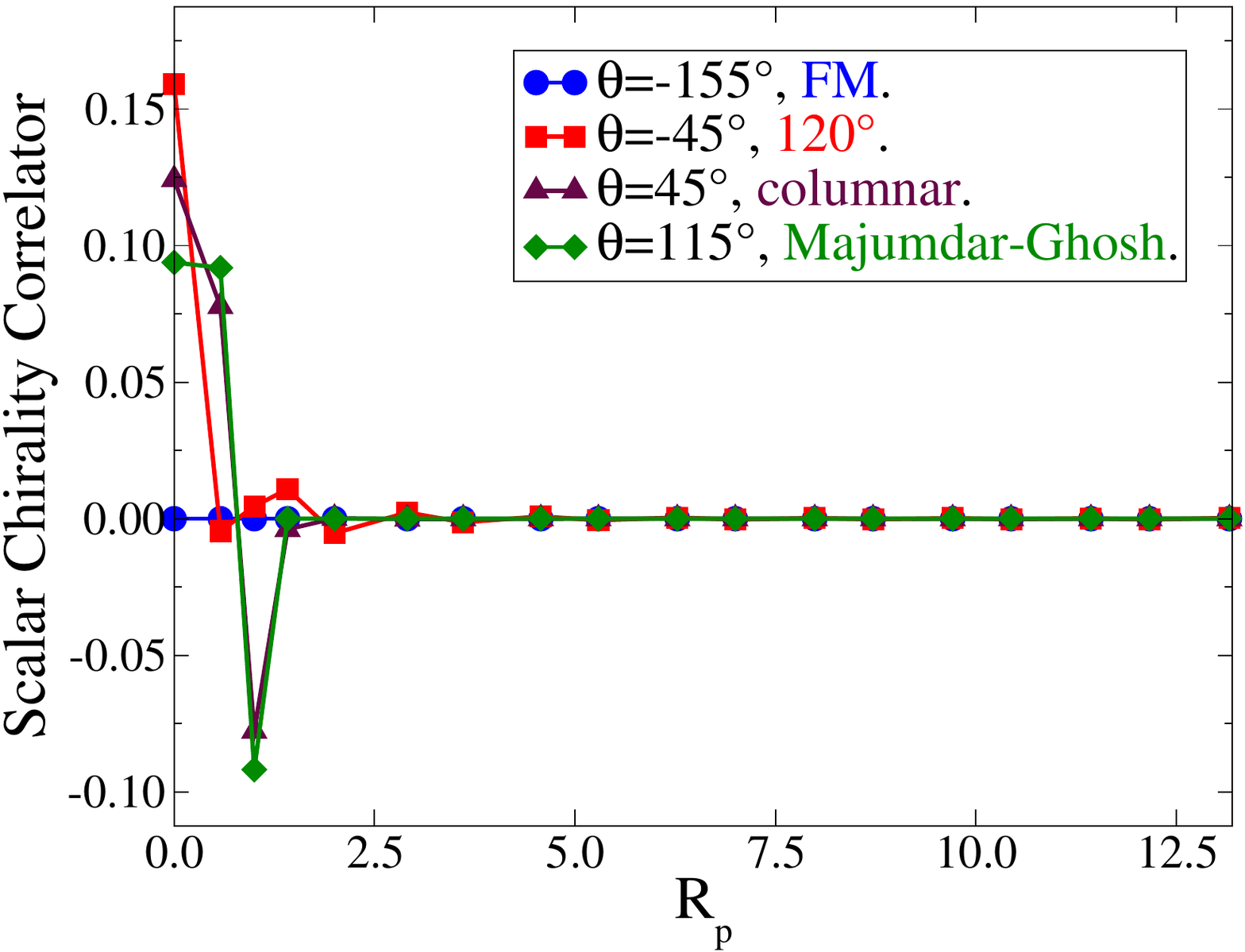}
    \caption{(Color online) Scalar chirality correlator, $O_t$,
      \eref{eq:CtCorr}, on a 30$\times$3 cylinder. $R_P$ is
      the distance between center of plaquettes in units of the lattice
      spacing. The path chosen to calculate these correlators is the
      same as the inset of \fref{fig:CrossCorr}. The rapid reduction of $O_t$ 
      to zero at long range indicates that the
      phases are non-chiral and planar.
      \label{fig:TripleCorr}}
  \end{center}
\end{figure}

A commonly considered
chiral order parameter for the THM is the triple product on a triangular plaquette,
known as the scalar chirality,
\begin{equation}
  \label{eq:triple}
  C_t[A,B,C] = {\bf S}_A . ( {\bf S}_B \times {\bf S}_C ) \; .
\end{equation}
The triple product operator breaks both P and T symmetries, and would acquire
different signs for different plaqeuttes according to their chirality.
A non-zero value of
$C_t[A,B,C]$ also implies that the spins are non-planar on that
plaquette. As a result a non-chiral and planar
system should acquire values close to zero for this triple product. 
Some studies\cite{Nishimori88,GroupChiral} predict
that the THM should be chiral in some circumstances (e.g. considering couplings higher
than two-body exchange interactions), while the
others\cite{Dagotto89,Jolicoeur90,Chubukov92,Deutscher93} strongly
suggest that the quantum fluctuations always select a planar spin
arrangement, so there is no chiral symmetry breaking.

It is important to note that,
the square of the triple product,
\eref{eq:triple}, on a single triangular plaquette is \emph{not} a good
order parameter to measure chirality, because it can be
shown,\cite{Wen89} similarly to the cross product, that for any spin-$1/2$ system,
\begin{equation}
  \label{eq:WWZ}
  \left({\bf S}_i \cdot ( {\bf S}_j \times {\bf S}_k)\right)^2 
  = -\frac{1}{16} M_{i,j,k}^2 + \frac{15}{64} \; .
\end{equation}
As a result $\langle C^2_t[A,B,C] \rangle$ on a plaquette is directly
related to the local magnetization and so is always non-zero, and
gives no indication of the chirality.

A diagnostic for the chirality is the correlator of the scalar chirality,
\begin{eqnarray}
  \label{eq:CtCorr}
  O_t(i,j,k;i',j',k') = \langle C_t[i,j,k] C_t[i^\prime,j^\prime,k^\prime] \rangle \; .
\end{eqnarray}
The results for the scalar chirality correlator are presented in
\fref{fig:TripleCorr}. The path here is same as the inset of
\fref{fig:CrossCorr}. All phases other than FM show
some short-range chiral correlations. However the rapid drop of $O_t$
to zero at long distance is a clear sign that all phases are non-chiral and planar.

\section{$120\,^{\circ}$ Order Parameter\label{sec:120}}

In the classical $120\,^{\circ}$ state on the triangular 
lattice every NNN bond is aligned ferromagnetically, while NN sites form
AFM bonds with uniform  expectation values, 
$\langle \vektor{S}_i . \vektor{S}_j \rangle = -1/8$. This state appears in a
semi-classical analysis.\cite{Jolicoeur90} 

The quantum analog of this classical $120^\circ$ state can be constructed by
positioning three spins at $120^\circ$ angles on the Bloch sphere,
forming a product state with long range order at 
wave vector $\vektor{Q} = (4 \pi / \sqrt{3},
4\pi/3)$, where the factor 2 arises from the rotation properties of spin-$1/2$
systems. The spin correlations
are $\langle \vektor{S}_i \cdot \vektor{S}_j \rangle = S_i S_j \cos 120^\circ = -1/8$ 
for each pair, which coincides with the classical value, as does the triple product 
$\vektor{S}_A \cdot (\vektor{S}_B \times \vektor{S}_C) = 0$. The plaquette magnetization 
$(\vektor{S}_A + \vektor{S}_B + \vektor{S}_C)^2 = 3/2$ is inherently non-classical.

\begin{figure}
  \begin{center}
    \includegraphics[width=\columnwidth]{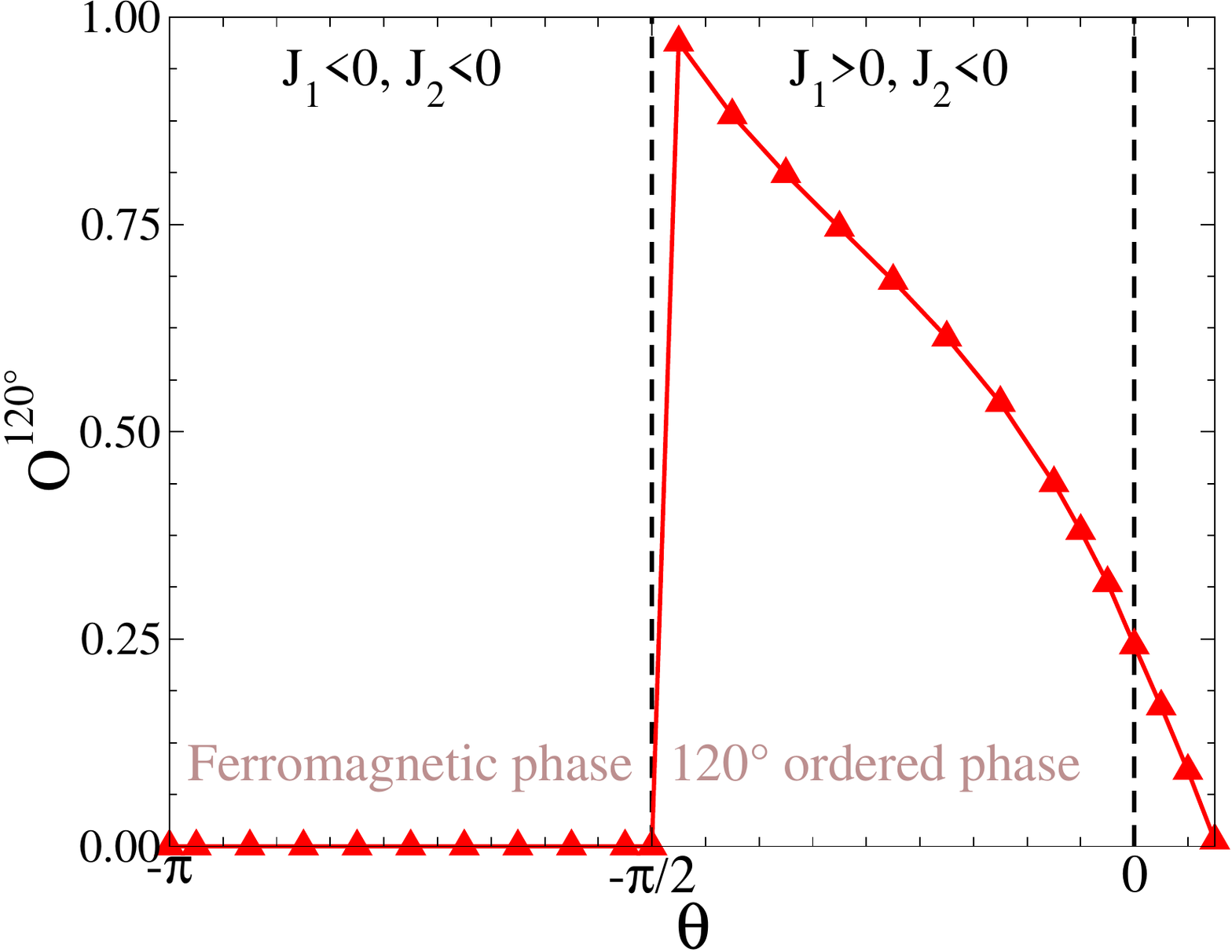}
    \caption{(Color online) Squared sublattice magnetization of $120^\circ$ state 
      on a $30\!\times\!3$ cylinder, $O^{120^\circ}$. 
      This order parameter correctly identifies the
      $120^\circ$ phase, consistent with the
      spin-spin correlations, \fref{fig:correlations}.
      \label{fig:120_order_param}}
  \end{center}
\end{figure}

A suitable order parameter
to detect this state is the squared sublattice magnetization of $120^\circ$ 
state and can be constructed as,
\begin{multline}
  E^{120^{\circ}} = \frac{1}{N_o} \sum_{i} \sum_{i^\prime} {\bf S}_i \cdot {\bf S}_{i^\prime}  \\
  \times \cos\left[\frac{4\pi}{\sqrt{3}}(x_i-x_{i^\prime})\right] \cos\left[\frac{4\pi}{3}(y_i-y_{i^\prime})\right] \;,
  \label{eq:120}
\end{multline}
where $N_o = N(N+4)/8$ is a normalization factor. 
$E^{120^\circ}$ will 
detect any state close to conventional $120^\circ$ order. 
DMRG results for the squared sublattice magnetization of $120^\circ$ state 
on a $30\!\times\!3$ cylinder, $O^{120^\circ} = \langle E^{120^\circ} \rangle$,
are shown in \fref{fig:120_order_param}. The region with non-zero values
of $O^{120^\circ}$ in \fref{fig:120_order_param} is consistent
with the $120^\circ$ phase region of \fref{fig:PhaseDiagram}, showing that this is a good
order parameter for the $120^\circ$ phase. The value of sublattice magnetization for the NN model,
$\sqrt{O^{120^\circ}(\theta=0)} \sim 49\%$ of the classical value, is comparable to previous calculations
on the 2D model, 50\% by ED,\cite{Bernu94} and 40\% by CCM.\cite{CCMSpinLiquid}

The $120^\circ$ order parameter is close to maximal in the limit $\theta \rightarrow -\pi/2$,
where the ground-state tends toward the quantum counterpart of the classical $120^{\circ}$ state.
This limit can be understood as fully saturated ferromagnetism
on each sublattice due to the large negative $J_2$;
a small positive $J_1$ then induces $120^\circ$ ordering between the sublattice.

\section{Phase Transitions and Critical Points}
\label{sec:critical}

In this section we pinpoint the location of the phase transitions
and their nature, determined from the magnetization, order parameters
and spin gaps.

\begin{figure}
  \begin{center}
    \includegraphics[width=\columnwidth]{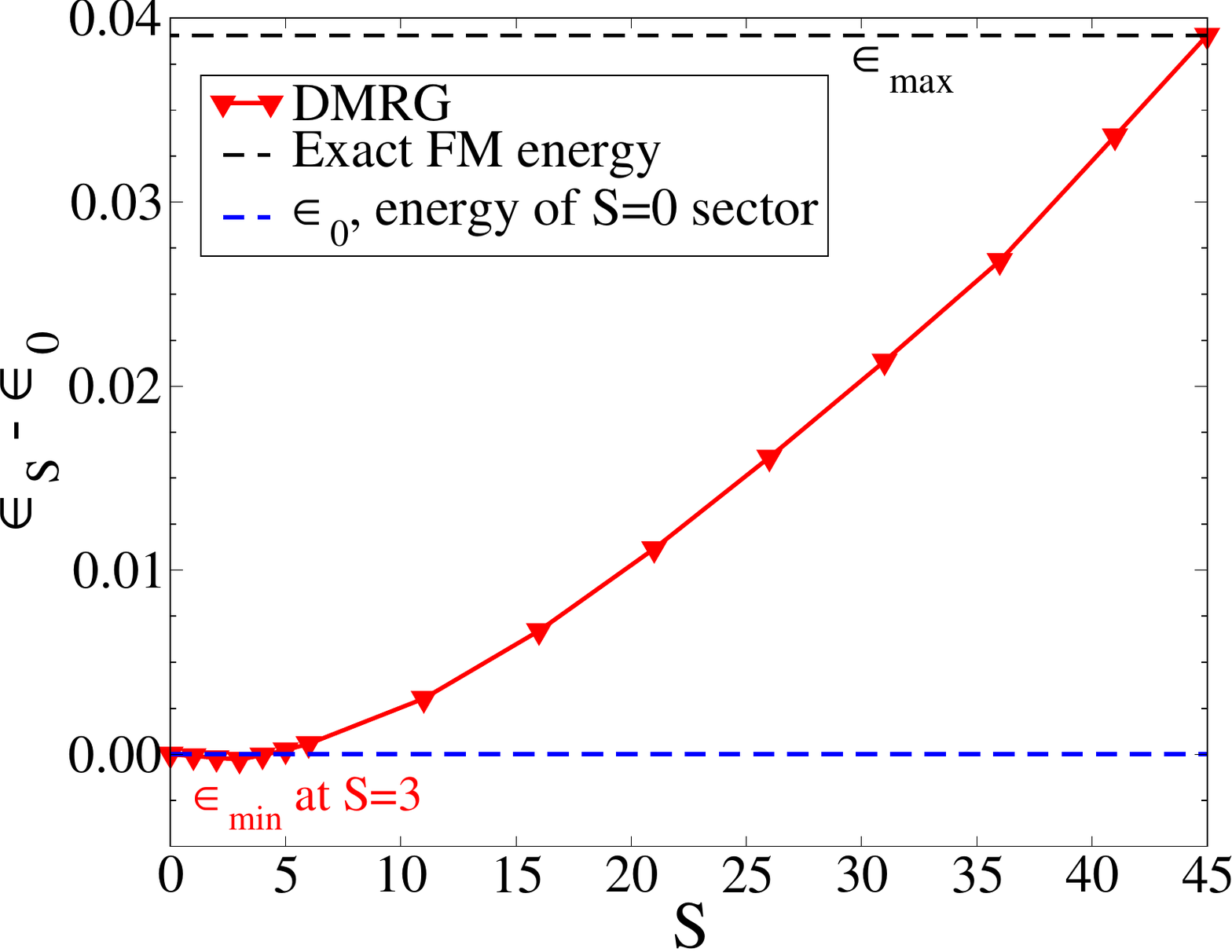}
    \caption{(Color online) Difference between energy per bond of
      partially polarized states and non-polarized state on a
      30$\times$3 cylinder at $\theta = 153\,^{\circ}$, near the
      critical point $\theta_c=152.0\,^{\circ}$. $S$ is the total spin,
      which is a good quantum number. The curve has minimum at a non-zero
      polarization, which indicates that there is a second-order phase
      transition close to this point.
      \label{fig:dE-S30.6}}
  \end{center}
\end{figure}

\begin{figure}
  \begin{center}
    \includegraphics[width=\columnwidth]{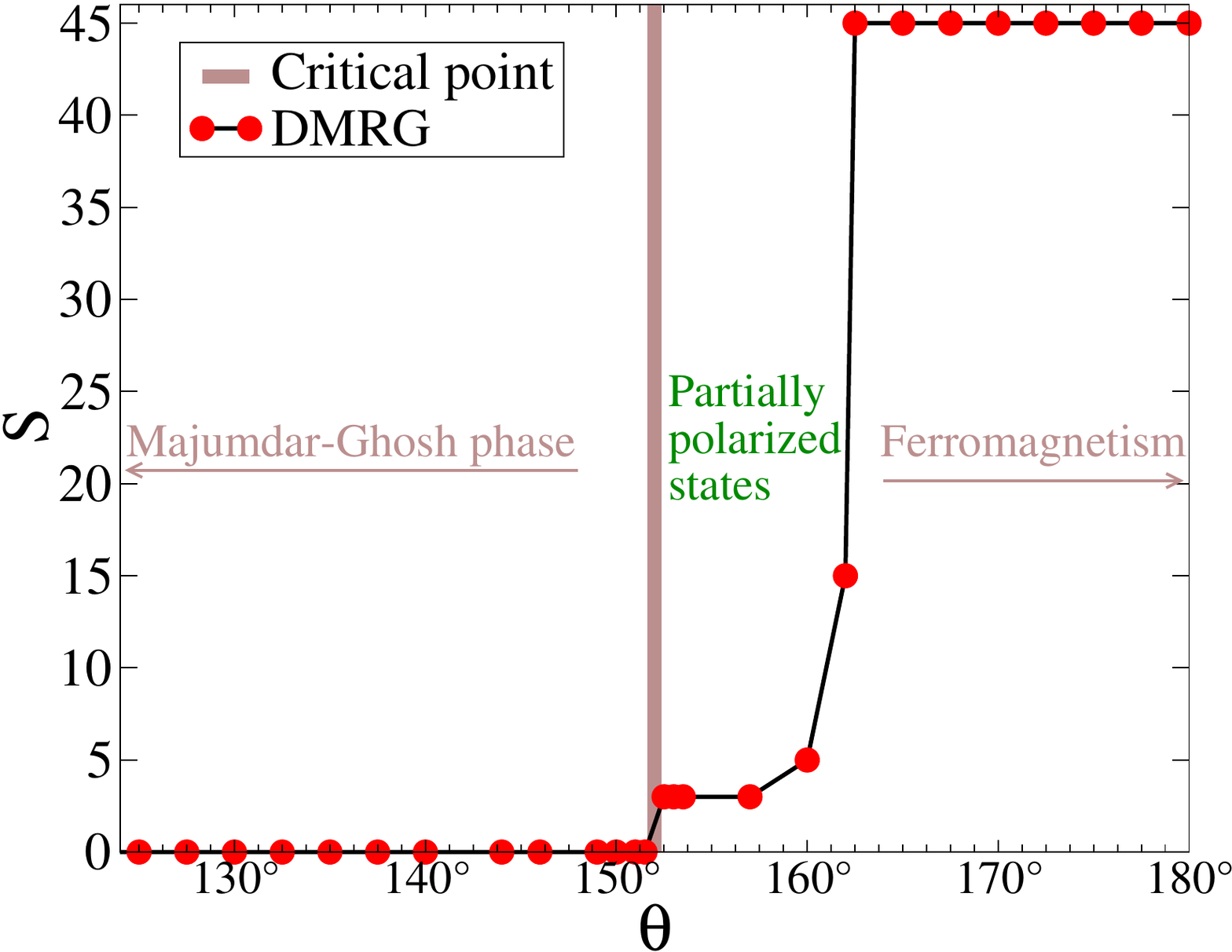}
    \caption{(Color online) Magnetization (ground-state spin
      polarization) around the critical point $\theta_c=152.0\,^{\circ}$
      on a 30$\times$3 cylinder. The onset of the phase transition
      is indicated with the brown stripe.
      \label{fig:S-theta}}
  \end{center}
\end{figure}


The point $\theta=-{\pi}/{2}$ 
marks rapid changes in many observables, consistent with a first order transition.
Indeed, since $J_1=0$ at this point, the ground-state consists of three uncoupled
sublattices, with ferromagnetic bonds within each sublattice, as discussed in \sref{sec:PhaseDiagram}. 
Therefore the ground-state
is $N/2$-fold degenerate, and hence the $120^\circ$ state and the fully polarized ferromagnet
coexist. 

To study the nature of the phase transition at
$\theta_c=152.0^{\circ}$ (NNN Majumdar-Ghosh to ferromagnet), 
we calculated the lowest-energy state in every possible total spin sector.
At points near the transition, we found a partially polarized ground-state.
For example, at $\theta=153^\circ$, shown in \fref{fig:dE-S30.6}, 
the ground-state for a $30\!\times\!3$ cylinder has total spin $S=3$.
This indicates a \emph{second-order} transition. 
We also calculated the ground-state magnetization around the critical
point, which is shown in \fref{fig:S-theta}. 
The obtained
transition point indicated by the brown stripe is consistent with
the correlation function results from \fref{fig:correlations}. 

The elementary excitations in the Majumdar-Ghosh chain are pairs of spin-$1/2$
solitons.\cite{Shastry81} In the NNN Majumdar-Ghosh phase of the 3-leg ladder, 
our numerical calculations show that the solitons in each sublattice
are pinned to each other, forming a dislocation line. Hence the elementary
excitations are a pairs of dislocations, with total spin $S=3$.

\begin{figure}
  \begin{center}
    \includegraphics[width=\columnwidth]{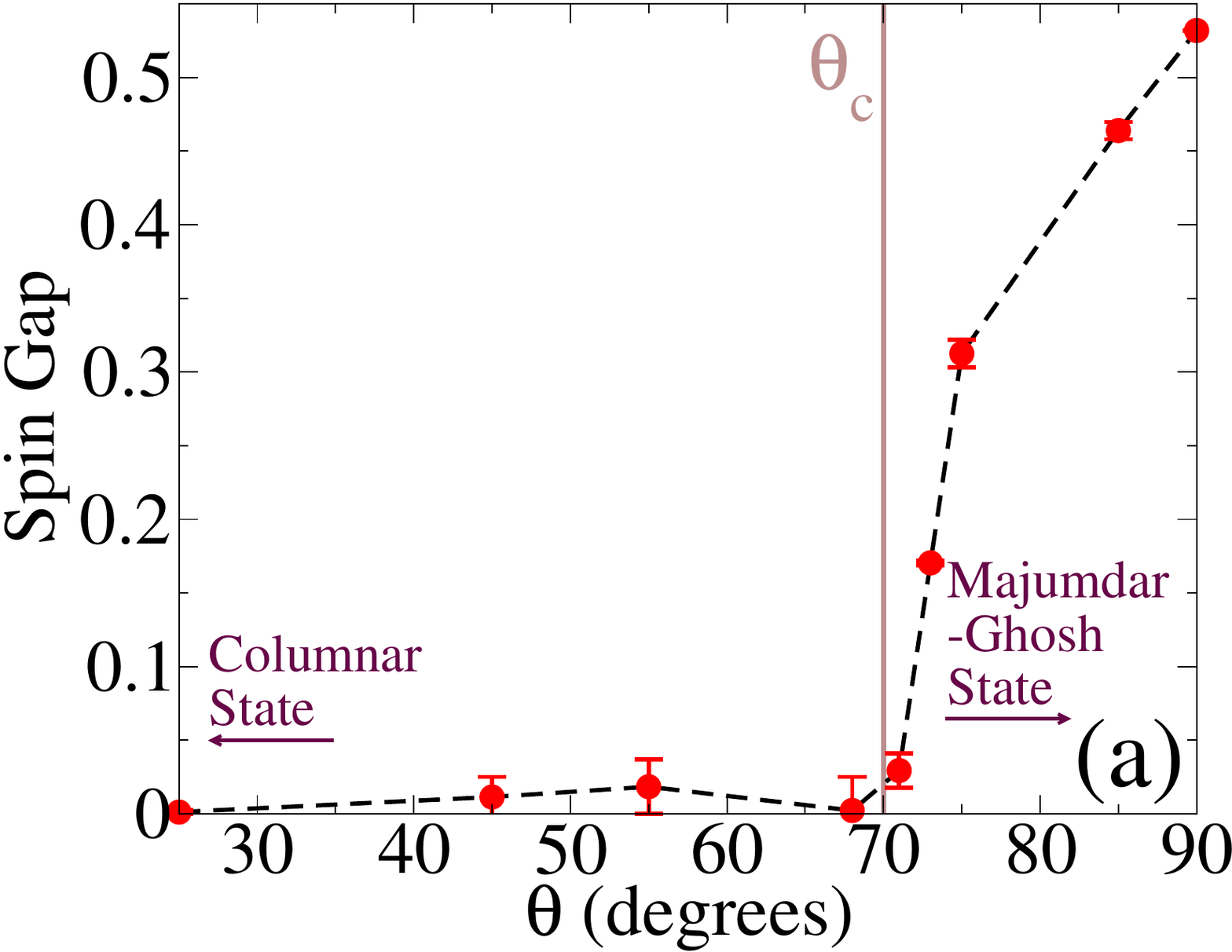}\\
    \includegraphics[width=\columnwidth]{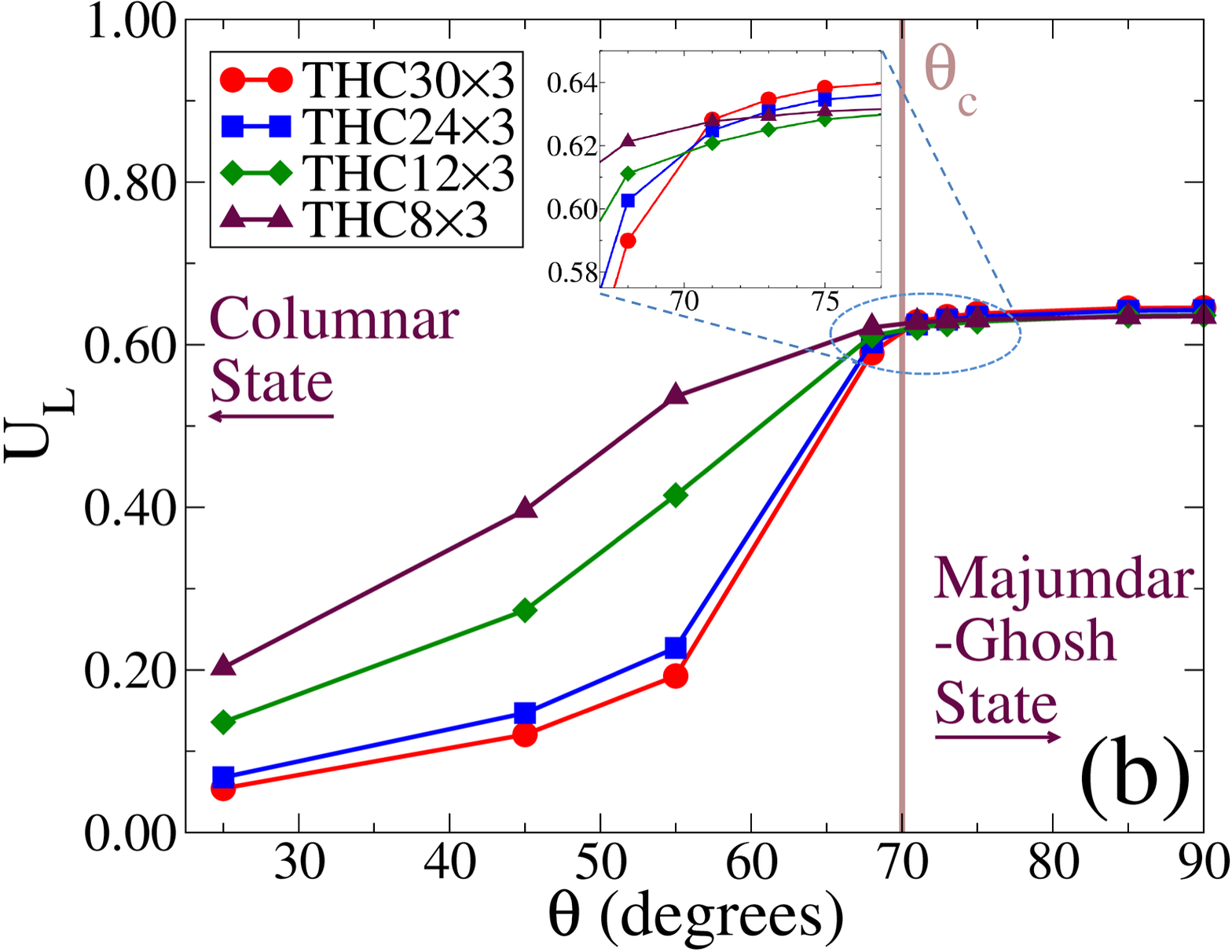}
    \caption{(Color online) (a) Spin gap extrapolated to the thermodynamic limit,
      (b) Binder cumulant of the dimer order parameter, \eref{eq:dimer}, 
      for the columnar to Majumdar-Ghosh phase transition.
      \label{fig:SpinGap_Binder}}
  \end{center}
\end{figure}

Accurately locating of the phase transition from the Majumdar-Ghosh state into the
gapless columnar state is more difficult. Deep in the columnar phase,
finite-size scaling of the spin gap is consistent with zero gap, as expected.
But the finite size
scaling is difficult to perform near the phase boundary because the
finite-size corrections in the two phases scale differently. Hence the spin gap
has fairly large error bars in this region, and the exact transition is difficult
to identify. One can use instead the dimer order parameter, defined for this model as,
\begin{equation}
 \label{eq:dimer}
 D_{(A)} = \frac{3}{N} \sum_{\langle\langle i,j \rangle\rangle \in A} (-1)^i \vektor{S}_i \cdot \vektor{S}_j \; ,
\end{equation}
where the sum is over all NNN spins in one sublattice.
However the dimer order parameter also contains large finite-size corrections.
A standard procedure (although not common in DMRG calculations) is to use higher
moments of the order parameter to cancel out low-order finite-size effects,
for example using the Binder cumulant,\cite{Binder81}
\begin{equation}
 U_L = 1 - \frac{\langle D^4 \rangle}{3\langle D^2 \rangle^2} \; .
\end{equation}
The spin gap and Binder cumulant of the dimer order parameter
are shown in \fref{fig:SpinGap_Binder}.
To obtain the spin gap, we firstly calculated the gap between 
$S=0$ and $S=1$ total spin sectors for finite-length cylinders. The gap was extrapolated 
to the thermodynamic limit, \fref{fig:SpinGap_Binder}(a), using the $L^{-3/2}$ scaling 
identified by Neuberger et al.,\cite{Neuberger89}
which produces a good fit except very close to the transition to the gapped
Majumdar-Ghosh phase.
The Binder cumulant,
\fref{fig:SpinGap_Binder}(b), shows the expected behavior, whereby
the value of the Binder cumulant at the phase transition is independent of the
lattice size (up to higher order corrections). The curves for $12\times 3$,
$24\times 3$ and $30 \times 3$ intersect quite closely, indicating that the transition is
in the vicinity of $\theta = 70.0^\circ$.

The columnar and $120^\circ$ phases are both gapless,
but we identify the location of the phase transition from the vanishing
of the short-range $O^{120^\circ}$ order parameter shown in \fref{fig:120_order_param},
giving the transition point as $\theta \simeq 6.5^\circ$.

\section{Conclusion\label{sec:conclusion}}

We have performed a comprehensive study of the phase diagram of the
triangular $J_1$-$J_2$ model on a 3-leg cylinder, using both
finite DMRG and iDMRG methods.
There are four distinct phases in this model. 
All phases are non-chiral and planar. 
The $120^\circ$ and columnar phases
are gapless with quasi-long-range correlations.
For large $J_2 > 0$, the geometry
of the ladder results in a Majumdar-Ghosh-like phase 
with long-range dimer order and a two-fold degenerate ground-state.
This phase is an effect of the restricted geometry, and only exists
for $L\times 3$ and $L \times 4$ cylinders.

Because we use a finite-width chain, the absence of $SU(2)$ symmetry-breaking 
magnetic ordering means that the long-range physics is rather different to the 2D model.
In the true 2D model, both the $120^\circ$ and columnar phases are expected 
to be $SU(2)$-broken long-range ordered. Thus on increasing the width of the cylinder,
we expect that the correlations will increase in magnitude and the gapless modes
arising from the 1D criticality
will evolve into Goldstone modes associated with the broken symmetry of the
order parameter. 

The short-range physics and structure of the phase diagram of the 3-leg ladder
agrees closely with known results for the 2D model, especially in the small $J_2$ region.
We find a transition from $120^\circ$ to columnar phases
at $\theta_c \simeq 6.5^\circ$, close to the classical value.
Further studies on larger width cylinders 
have clarified that between the $120^\circ$ and columnar state there is
a spin liquid region,\cite{InPrep} consistent with the recent results of 
quantum Monte Carlo calculations.\cite{QMCSpinLiquid1,QMCSpinLiquid2} 

The boundary between 1D and 2D physics in this model is rather rich, and 
this suggests that the physics arising from restricting geometry to 
finite-width ladders presents a fruitful direction for future 
investigation, and may explain some novel properties of molecular
solids.\cite{FraxedasBook}


\begin{acknowledgments}

We thank C. Janani and F. Zhan for useful discussions. 
This work has been supported by the Australian Research
Council (ARC) Centre of Excellence for Engineered Quantum Systems, grant CE110001013. 
BJP is supported by the ARC under grant FT130100161.

After completion of this work, we became aware of some related results.\cite{WhiteJ1J2,Hu15}

\end{acknowledgments}


\cleardoublepage

\end{document}